\documentclass[a4paper,11pt]{article}
\usepackage{jheppub} 
\usepackage{lineno}
\usepackage{float}
\usepackage{comment}
\usepackage{tikz}
\usepackage{amsmath}
\usetikzlibrary{decorations.pathmorphing, decorations.markings, arrows.meta}

\tikzset{
  quark/.style={
    thick,
    postaction={decorate},
    decoration={
      markings,
      mark=at position 0.55 with {\arrow[thick]{Latex}}
    }
  },
  antiquark/.style={
    thick,
    postaction={decorate},
    decoration={
      markings,
      mark=at position 0.55 with {\arrowreversed[thick]{Latex}}
    }
  },
  gluon/.style={
    thick,
    decorate,
    decoration={
      coil,
      aspect=0.3,
      segment length=5pt,
      amplitude=4pt
    }
  }
}

\title{\boldmath Gluon radiation from a QCD antenna with realistic parton-medium interactions}

\author[a]{Carlota Andr\'es}
\author[b,c]{, Liliana Apolinário}
\author[d]{, N\'estor Armesto}
\author[b,c]{, Andr\'e Cordeiro}
\author[a,d]{, Fabio~Dom\'{\i}nguez}
\author[b,d]{, Pablo Guerrero-Rodr\'iguez}
\author[b,c]{, Jos\'e Guilherme Milhano}
\affiliation[a]{CPHT, CNRS, \'Ecole polytechnique, Institut Polytechnique de Paris, 91120 Palaiseau, France}
\affiliation[b]{LIP - Laborat\'{o}rio de Instrumentação e F\'{i}sica Experimental de Part\'{i}culas,\\
Av. Prof. Gama Pinto, 2, 1649-003, Lisbon, Portugal}
\affiliation[c]{Departamento de Física, Instituto Superior Técnico, Universidade de Lisboa,\\
Av. Rovisco Pais 1, 1049-001 Lisbon, Portugal}
\affiliation[d]{Instituto Galego de F\'{\i}sica de Altas Enerx\'{\i}as IGFAE,
Universidade de Santiago de Compostela,\\ 15782 Santiago de Compostela, Galicia-Spain}

\emailAdd{carlota.andres@polytechnique.edu}
\emailAdd{liliana@lip.pt}
\emailAdd{nestor.armesto@usc.es}
\emailAdd{andre.cordeiro@tecnico.ulisboa.pt}
\emailAdd{fabio.dominguez@usc.es}
\emailAdd{pablo.guerrero@usc.es} 
\emailAdd{gmilhano@lip.pt} 

\abstract{The spectrum of coherent gluon radiation from a quark-antiquark pair undergoing multiple scatterings within a colored medium is central for understanding in-medium parton cascades.
However, current efforts are constrained by reliance on a number of approximations, such as the harmonic oscillator approximation, that are only valid within limited regions of phase space.
In this paper, we circumvent this problem by expressing the full in-medium gluon emission spectrum as a set of differential equations that can be solved numerically. This formalism, previously applied to the case of medium-induced radiation off a single color charge, allows to resum medium interactions to all orders while employing realistic scattering models. The resulting angle and energy distributions of emitted gluons serve to illustrate the breakdown of color coherence across the entire accessible phase-space, and constitute a definite step towards a higher-precision description of jet observables.}

\usepackage{soul}

\newcommand{\diff}{{\rm d}} 

\newcommand{\bm}[1]{\boldsymbol{#1}}
\newcommand{\bvec}[1]{\boldsymbol{#1}}

\newcommand{\Smed}{\mathcal{S}_{\rm med}}
\newcommand{\kappabar}{{\bar{\kappa}}}
\newcommand{\dn}{\delta n}

\newcommand{\omegaC}{\bar{\omega}_c}
\newcommand{\thetaC}{\bar{\theta}_c}
\newcommand{\RC}{R_c} 
\newcommand{\rk}{r_\kappa} 
\newcommand{\nk}{\theta} 
\newcommand{\nkbar}{\bar{\theta}} 

\newcommand{\md}{m_\text{D}}

\newcommand{\psiAB}{\psi^{\,}_\text{AB}}            
\newcommand{\psiABavg}{\tilde{\psi}^{\,}_\text{AB}}

\newcommand{\gAB}{h}
\newcommand{\GAB}{H}

\begin{document}
\maketitle
\flushbottom

\section{Introduction}
\label{sec:intro}

It is by now well established that jets are among the most powerful probes of the quark–gluon plasma (QGP) formed in ultra-relativistic heavy-ion collisions \cite{Busza:2018rrf,Harris:2023tti,Wang:2025lct,Cao:2020wlm,Apolinario:2022vzg,Cunqueiro:2021wls, CMS:2024krd,ALICE:2022wpn}. For many years, jet quenching studies focused primarily on  energy loss calculations, which provided a natural interpretation of early experimental measurements such as di-hadron correlations  and the hadron nuclear modification factor $R_{\rm AA}$ \cite{PHENIX:2001hpc,STAR:2002svs,STAR:2003pjh}. More recently, there has been a growing interest in the modification of the internal structure of jets as a probe of both the medium properties and the dynamics of in-medium jet evolution, see for instance~\cite{Apolinario:2024equ,Apolinario:2022vzg,Cunqueiro:2021wls,Connors:2017ptx,CMS:2024krd}.

One of the key concepts in understanding jet modification in a QCD medium is color decoherence. Prior to hadronization, jets in vacuum evolve in a color-coherent manner: the conservation of the total color charge of the initiating parton leads to angular ordering, where large-angle emissions must occur early in the shower and are sensitive only to the total color charge, while later emissions at smaller angles can resolve individual parton color charges of the branching partons. This  picture is modified when the jet evolves inside a colored medium, where  multiple interactions between its constituents and the QGP break the color coherence of the jet. This loss of coherence not only induces additional medium-induced radiation, but also allows a larger fraction of the energy to be distributed at large angles with respect to the jet axis.

The first calculations of medium-induced radiation incorporating this phenomenon \cite{Mehtar-Tani:2010ebp,Mehtar-Tani:2011hma,Casalderrey-Solana:2011ule,Mehtar-Tani:2012mfa,Casalderrey-Solana:2012evi} are framed in terms of emissions off a quark-antiquark pair in a definite overall color state, commonly referred to as antenna. In this setup, the quark and antiquark in the antenna are taken to be extremely energetic and to follow straight-line (eikonal) trajectories. The emitted gluons carry much smaller energy and are treated in the soft limit, in which recoil effects on the emitters can be neglected. In vacuum, the suppression of radiation at large angles arises from the destructive interference between emissions off the two legs of the antenna, which can only occur if the overall color state of the system is preserved. In the presence of a medium, multiple color exchanges between the antenna constituents and the medium suppress the interference term, thus enhancing the probability of emissions at large angles compared to the vacuum case. The rate at which this color decoherence occurs depends on both the properties of the medium and the opening angle of the antenna.

These effects must therefore be accounted for the correct description of the modification of jets while traversing a QCD medium. Color decoherence not only alters the internal structure of jets, but also affects their energy loss, since the loss of coherence effectively increases the number of independent radiation sources. In particular, early splittings can generate color antennas that may fully decohere in the medium and subsequently radiate independently, enhancing large-angle emissions. As a result, decoherence signatures have been actively searched for in several experimental analyses \cite{CMS:2017qlm,ALargeIonColliderExperiment:2021mqf,ATLAS:2022vii,ATLAS:2025svn}, and their dynamics have been incorporated into phenomenological studies \cite{Caucal:2019uvr,Caucal:2020xad,Mehtar-Tani:2021fud,Mehtar-Tani:2024jtd,Apolinario:2026hff}, where jet energy loss accounts for the possibility of multiple sources within the same jet.  Furthermore, different effective prescriptions for coherence inspired in such calculations have been incorporated in some jet quenching Monte Carlo  simulators~\cite{ Hulcher:2017cpt,Caucal:2018dla,Caucal:2019uvr,Casalderrey-Solana:2019ubu,Zapp:2026cqf}.

Despite those phenomenological efforts to include color decoherence in the description of jet quenching data, it must be acknowledged that the main hurdle in improving the accuracy of those descriptions are the approximations underlying the original antenna calculations. In particular, two main assumptions in the original antenna calculations need to be relaxed to achieve a more realistic treatment of color decoherence in medium-modified jets: the extremely high-energy approximation for the quark and antiquark, which effectively enforces instantaneous antenna formation and neglects recoil from medium scatterings, and the simplified treatment of jet–medium interactions, which are taken either in the first opacity limit or in the multiple soft scattering approximation. Relaxing the former is particularly challenging, as  accounting for the dynamics of antenna formation significantly increases the complexity of the calculation. Efforts in this direction can be broadly grouped into two categories:  calculations attempting to understand the in-medium splittings in which the antenna would be formed~\cite{Blaizot:2012fh,Apolinario:2012vy,Apolinario:2014csa,Dominguez:2019ges,Isaksen:2023nlr,Andres:2026qrt,Leitao:2026fgh,Barata:2021byj}, and studies aiming at quantifying the impact of a finite formation time on the decoherence process~\cite{Abreu:2024wka}\footnote{We further note recent progress in a complementary direction, where transverse momentum broadening in antennas has been reformulated in terms of an open quantum system approach~\cite{Arleo:2026urv}. Simultaneously,~\cite{Barata:2026icn} developed a framework that employs quantum simulation techniques to compute multi-particle processes in a dense media}. Improving the description of in-medium antennas through a more realistic treatment of multiple scatterings is the main focus of this work.

Most jet quenching perturbative calculations have suffered from the same shortcomings over the years. Even though formal expressions resumming multiple scatterings to all orders are known, complete computations which can be used for numerical evaluation are possible only under restrictive approximations. Two different approaches have been commonly used to obtain approximate solutions: the single scattering approach (or first order in an opacity expansion) and the resummation of multiple soft scatterings (also known as the harmonic oscillator approximation). While both approaches have been useful to understand qualitative features of medium-induced emissions in different regimes, they fail at giving a reliable quantitative description over a large region of phase space: the former gives the right behavior at large energies and large transverse momentum absent in the latter, but it misses the suppression effect of multiple scatterings responsible of maintaining unitarity in certain regions of phase space \cite{Andres:2020kfg}. In recent years, an effort to combine the advantages of these two approaches has been developed in the form of the improved opacity expansion (IOE)~\cite{Mehtar-Tani:2019tvy,Mehtar-Tani:2019ygg,Barata:2021wuf} where the multiple soft scattering approach is taken as the zeroth order of an expansion in terms of \emph{hard} scatterings. It certainly achieves much better results than any of the two approaches individually, although at the cost of introducing some arbitrary matching scales. Even though the IOE was originally proposed for the single emission spectrum from a single source, it was recently used to calculate soft emissions from an antenna~\cite{Kuzmin:2025fyu}.

In this manuscript, we follow the approach of~\cite{Andres:2020vxs} where a full resummation of multiple scatterings was performed for realistic probe-medium interactions without any further approximations in the context of soft emissions off a single parton. In that approach, the different factors entering the evaluation of the in-medium emission spectrum are expressed as solutions of integro-differential equations which are then treated numerically (see also \cite{Caron-Huot:2010qjx}). Extending this formulation to the antenna case requires overcoming several obstacles, most notably the lack of symmetry in analytically performing some of the angular integrations. In this work, we present results for the emission of soft gluons off a quark-antiquark antenna traversing a medium, including full resummation of multiple interactions with a realistic collision rate describing the interactions between the partons and the medium. We consider two different parton-medium interaction rates (Yukawa~\cite{Wang:1992qdg} and hard thermal loop~\cite{Aurenche:2002pd}), thereby showing the flexibility of the approach.

This paper is organized as follows: In section~\ref{sec:qcd-antenna} we review the main concepts underlying  the propagation  and radiation off a QCD antenna in a colored medium. In section~\ref{sec:diff-eqs}, we present the formalism for the calculation of the in-medium gluon spectrum, formulated in terms of a set of coupled integro-differential equations. We specify the initial conditions, discuss different parton–medium interaction models, and  provide analytical results in the single-scattering limit. In section~\ref{sec:results}, we present and discuss our numerical results. Finally, section~\ref{sec:conclusion} contains our conclusions. Additional technical details and supplementary numerical results are provided in the appendices.

\section{QCD antenna in a colored medium}
\label{sec:qcd-antenna}

\begin{figure}[ht]
\centering
\begin{minipage}{0.48\textwidth}
    \centering
    \usetikzlibrary{decorations.pathmorphing, decorations.markings, arrows.meta}

\tikzset{
  quark/.style={
    thick,
    postaction={decorate},
    decoration={
      markings,
      mark=at position 0.55 with {\arrow[thick]{Latex}}
    }
  },
  antiquark/.style={
    thick,
    postaction={decorate},
    decoration={
      markings,
      mark=at position 0.55 with {\arrowreversed[thick]{Latex}}
    }
  },
  gluon/.style={
    thick,
    decorate,
    decoration={
      coil,
      aspect=0.3,
      segment length=5pt,
      amplitude=4pt
    }
  }
}

\begin{tikzpicture}[x=0.75cm, y=0.75cm, font=\small]

  \coordinate (O)    at (0.0,  0.0);   
  \coordinate (Vg)   at (2.5,  1.1);   
  \coordinate (Qout) at (5.0,  2.2);   
  \coordinate (Qbar) at (5.0, -1.6);   
  \coordinate (Gout) at (4.8,  4.0);   

  \filldraw[fill=gray!25, draw=black, thick] (O) circle (9pt);

  \draw[quark] (O)  -- (Vg);
  \draw[quark] (Vg) -- (Qout);

  \draw[antiquark] (O) -- (Qbar);

  \draw[gluon] (Gout) -- (Vg);

  \filldraw[black] (Vg) circle (2.5pt);

  \node[right=5pt] at (Qout)
    {$q\;\bigl(E = p^{+},\,\boldsymbol{p}\bigr)$};

  \node[right=5pt] at (Qbar)
    {$\bar{q}\;\bigl(\bar{E} = \bar{p}^{+},\,\bar{\boldsymbol{p}}\bigr)$};

  \node[right=5pt] at (Gout)
    {$g\;\bigl(\omega = k^{+},\,\boldsymbol{k}\bigr)$};

\end{tikzpicture}
\end{minipage}
\hfill
\begin{minipage}{0.48\textwidth}
    \centering
    \usetikzlibrary{arrows}
\usetikzlibrary{arrows.meta}
\begin{tikzpicture}[scale=1.0
                    ,>={stealth[fill=black]}
                    , thick]
    \draw[->, dotted] (0,-2.2) -- (0,3) node[above] {$p_x/p^{+}$};
    \draw[->, dotted] (-0.4,-0.25) -- (2.5,-0.25) node[right] {$p_y/p^{+}$};
    \coordinate  (Q)     at (0.0, +2.2);
    \coordinate  (QBAR)  at (0.0, -1.6);
    \coordinate  (GLUON) at (125.0:+3.5cm);
    \node[black, right] at (Q)     {$\boldsymbol{p}/E$};
    \node[black, right] at (QBAR)  {$\boldsymbol{\bar{p}}/\bar{E}$};
    \node[black, above] at (GLUON) {$\boldsymbol{k}/\omega$};
    \draw[fill, black] (Q)     circle (1.5pt);
    \draw[fill, black] (QBAR)  circle (1.5pt);
    \draw[fill, black] (GLUON) circle (1.5pt);
    \draw[->, thick, black, shorten >=3pt, shorten <=3pt] (Q) -- (GLUON) 
        node[pos=0.45, below, sloped]        {$\boldsymbol{\theta} = \boldsymbol{\kappa}/\omega$};
    \draw[->, thick, black, shorten >=3pt, shorten <=3pt] (QBAR) -- (GLUON) 
        node[midway, below, xshift=-18pt]         {$\boldsymbol{\bar{\theta}} = \boldsymbol{\bar{\kappa}}/\omega$};
    \draw[->, thick, black, shorten >=4pt, shorten <=3pt] (Q) -- (QBAR) 
        node[pos=0.35, right]        {$\boldsymbol{\delta n}= \boldsymbol{\delta\kappa}/\omega$};
        
    \node[anchor=north east] at (3.8,3.2) {(Transverse view)};
\end{tikzpicture}
\end{minipage}
\caption{ Left:  Schematic representation of the antenna configuration, indicating the momenta of the three legs. Right: Illustration of the kinematic variables used to describe the transverse degrees of freedom for the final state gluon.}
\label{fig:antenna}
\end{figure}
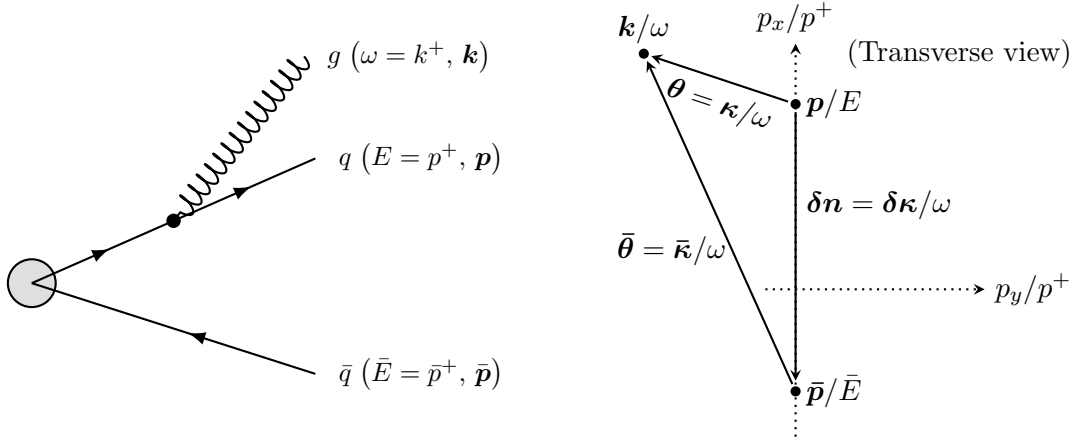

We start by reviewing  the general formalism used to compute the gluon emission spectrum of a QCD antenna propagating through a colored medium and undergoing and arbitrary number of scatterings \cite{Casalderrey-Solana:2011ule,Mehtar-Tani:2011lic,Mehtar-Tani:2012mfa}. It is important to emphasize that this calculation is performed in a high-energy approximation with the following standard assumptions:
\begin{itemize}
    \item Parton-medium interactions are modelled as transverse momentum and color exchanges with a medium. Following the ideas in the Gyulassy-Wang~\mbox{\cite{Gyulassy:1993hr}} and McLerran-Venugopalan~\mbox{\cite{McLerran:1993ka}} models, the medium is considered as transversely homogeneous and isotropic, and described by a classical background field with color correlators instantaneous in light-cone\footnote{%
    Throughout this manuscript, we use light-cone coordinates, where $v^\pm = (v^0 \pm v^3)/\sqrt{2}$ and $\bm{v}=(v^1,v^2)$.
    Transverse integrations are written $\int_{\bm{q}} = \int \diff^2\bm{q}/(2\pi)^2 $ in momentum space and $\int_{\bm{x}} = \int \diff^2 \bm{x}$ in coordinate space.
    } time $t=x^+$, which simplifies the averages over medium color configurations.
    \item The transverse momentum exchanges between the antenna and the medium constituents are assumed to be dominated by a characteristic medium scale, namely the temperature in the case of a thermal medium or the saturation scale in the case of a dense hadron.  Transverse momentum exchanges larger than that scale are suppressed.
    \item The antenna consists of an ultra-relativistic quark-antiquark pair %
    whose  longitudinal light-cone momentum $p^+$ is not altered by the gluon emission. Accordingly, the quark and antiquark kinematics are given in light-cone coordinates as $(p^+\equiv E, \bm{p})$ and $(\bar{p}^+ \equiv \bar{E}, \bar{\bm{p}})$ respectively, while the gluon momentum is specified by $(k^+\equiv \omega,\bm{k})$, see figure~\ref{fig:antenna} (left). Similarly to calculations for the factorization of vacuum-like antenna emissions~\cite{Dokshitzer:1991wu,Ellis:1996mzs,Campbell:2022hdd}, the gluon is assumed to be soft, i.e.  $\omega \ll E, \bar{E}$.\footnote{%
    We note that calculations of finite-energy gluon emissions off a single parton  can be found in~\cite{Blaizot:2012fh,Apolinario:2012vy,Apolinario:2014csa,Isaksen:2023nlr, Andres:2026qrt}.
    }
    \item The antenna formation is assumed to be instantaneous. As a consequence, it is not resolved by the medium and it does not overlap in time with the gluon emission\footnote{%
    Calculations including finite formation time under the restrictive semi-hard approximation can be found  in~\cite{Dominguez:2019ges,Isaksen:2020npj,Abreu:2024wka}, and  overlapping formation times were considered in~\cite{Casalderrey-Solana:2015bww,Arnold:2015qya,Arnold:2016kek,Arnold:2016mth,Arnold:2016jnq,Arnold:2020uzm,Arnold:2021pin,Arnold:2022epx,Arnold:2022fku,Arnold:2022mby,Arnold:2023qwi,Arnold:2024whj}.}. 
    \item The quark and antiquark propagators are taken in the eikonal approximation (probing the medium along straight lines), while the gluon propagator includes sub-eikonal corrections (probing the medium along all arbitrary paths). This follows early calculations for in-medium soft gluon radiation from a hard parton~\cite{Baier:1996kr,Zakharov:1996fv,Wiedemann:2000za,Gyulassy:2000er} and the original in-medium antenna calculations~\cite{Mehtar-Tani:2010ebp,Mehtar-Tani:2011hma,Casalderrey-Solana:2011ule,Mehtar-Tani:2012mfa,Casalderrey-Solana:2012evi}
\end{itemize}

In general, the emission rate for soft gluons off a quark-antiquark antenna can be split into color singlet and octet contributions~\cite{Mehtar-Tani:2011hma,Mehtar-Tani:2012mfa}, as follows 
\begin{equation}
    \omega\dfrac{\diff^3 I}{\diff\omega \diff^2\bvec{k}} 
    = 
    \dfrac{\alpha_s}{(2\pi)^2 \omega^2} 
    \left( C_F \mathcal{R}_\text{sing} + C_A \mathcal{J} \right) 
    ,
\label{eq:emission-rate}
\end{equation}
where the singlet and octet contributions are proportional to the color SU($N_c$) invariants~$C_F=\tfrac{N_c^2-1}{2N_c}$ and $C_A = N_c$ respectively. These are both functions of the transverse momenta of the gluon relative to the quark $\bm{\kappa} \equiv \bm{k} - \omega \bm{p} / E$, and to the antiquark $\bm{\bar{\kappa}} \equiv \bm{k} - \omega \bm{\bar{p}} / \bar{E}$. These obey:
\begin{equation}
    \mathcal{R}_\text{sing}(\bm{\kappa}, \bm{\kappabar}) = 
    \mathcal{R}(\bm{\kappa}) 
    + \mathcal{R}(\bm{\kappabar}) 
    - 2\mathcal{J}(\bm{\kappa}, \bm{\kappabar})
    \,,
    \text{ with }
    \mathcal{R}(\bm{\kappa}) = 
    \mathcal{J}(\bm{\kappa}, \bm{\kappa})
    \,.
\end{equation}
The emission rate~\eqref{eq:emission-rate} thus reduces to the computation of the function $\mathcal{J}(\bm{\kappa}, \bm{\kappabar})$, which encodes the interference between gluon radiation from the quark and antiquark. 
This interference term can be written in terms of the momentum broadening factor $\mathcal{P}$, the emission kernel $\mathcal{K}$, and the antenna decoherence parameter $\Delta_\text{med}$, as given in \cite{Casalderrey-Solana:2011ule,Mehtar-Tani:2012mfa}:
\begin{equation}
\begin{aligned}
	\mathcal{J}(\bm{\kappa},\bm{\kappabar}) 
    =
	\text{Re}	
	\int^\infty_0 \diff t \,
    &
	\int^{\infty}_t \diff\bar{t}
	\int_{\bm{\ell}\bm{q}}
    e^{ i t \tfrac{\bm{\bar{\ell}}^2 - \bm{\ell}^2}{2\omega} }
	(\bm{q} \cdot 
    \bm{\bar{\ell}})
    \\ 
    &
    \times\,
    \mathcal{P}(\bm{\kappa} - \bm{q}; \infty, \bar{t})
	\,\mathcal{K}(\bm{q}, \bar{t}; \bm{\ell}, t)
	\,[1-\Delta_{\rm med}(t)]    
    + (\bm{\kappa} \leftrightarrow \bm{\kappabar})
    \,.
\end{aligned}
\label{eq:J-term}
\end{equation}
Here and throughout the manuscript, for any integration variable $\bm{\ell}$, we write $\bm{\bar{\ell}} \equiv \bm{\ell} - \omega\bm{\dn}$, where 
\begin{equation}
    \bm{\dn} \equiv (\bm{\kappa} - \bm{\kappabar})/\omega = \bm{\bar{p}}/\bar{E} - \bm{p}/E
    \,,
\label{eq:antenna_angle}
\end{equation}
is a proxy for the antenna opening angle. The relation between $\bm{\dn}$ and the relative transverse momentum vectors is depicted in the right panel of  figure~\mbox{\ref{fig:antenna}}, together with the vectors $\bm{\nk}\equiv\bm{\kappa}/\omega$ ($\bm{\bar{\nk}}\equiv\bm{\bar{\kappa}}/\omega$), whose magnitudes measure the opening angle between the gluon and, respectively, the quark and antiquark.

Within the assumptions summarized above, the functions $\mathcal{P}$, $\mathcal{K}$, and $\Delta_\text{med}$ entering~\eqref{eq:J-term} can be obtained by resumming  an arbitrary number of parton-medium scatterings. One is therefore required to specify a model for the parton–medium interaction, which can in general be characterized in terms of the linear density of scattering centres $n(t)$ and the dipole cross-section $\sigma(\bm{r})$, itself determined by the scattering rate $V(\bm{q}^2)$
\begin{equation}
    \sigma(\bm{r}) = \int_{\bm{q}} V(\bm{q}^2) 
    \left(1-e^{i\bm{q}\cdot\bm{r}}\right)
    \,.
\label{eq:dipole}
\end{equation}
No further assumptions are made regarding the scattering rate, although the short-distance nature of Coulomb interactions implies the asymptotic behavior $V(\bm{q}^2\gg \mu^2)\sim \mu^2/\bm{q}^4$, where $\mu$ is set by the thermal scale of the medium.

Within this formalism, the momentum broadening factor, emission kernel, and decoherence parameter admit the following expressions:
\begin{align}
    \mathcal{P}(\bm{\kappa}-\bm{q}, \tau;  \bar{t}) &=
    \int_{\bm{z}} e^{-i\bm{z}\cdot(\bm{\kappa}-\bm{q})}
    \exp\left\{
    -\dfrac{1}{2}\int^{\tau}_{\bar{t}} \diff s \, n(s)\, \sigma(\bm{z}) 
    \right\}
    \,,
    \label{eq:broadening}
    \\
    \mathcal{K}(\bm{q}, \bar{t}; \bm{\ell}, t) &=
    \int_{\bm{xy}} e^{-i(\bm{q}\cdot\bm{y}-\bm{\ell}\cdot\bm{x})}
    \int^{\bm{r}({\bar{t}})=\bm{y}}_{\bm{r}(t)=\bm{x}} 
    \mathcal{D}\bm{r}
    \exp\left\{
    i\int^{\bar{t}}_{t} \diff s 
    \, \dfrac{\omega}{2}\dot{\bm{r}}(s)^2 + \dfrac{i}{2} n(s) \sigma(\bm{r})
    \right\}
    \,,
    \label{eq:kernel}
    \\
    \mathcal{S}_\text{med}(t) &\equiv 1 - \Delta_\text{med}(t)
    =
    \exp\left\{
    -\dfrac{1}{2}\int^{t}_{0} \diff s \, n(s) \sigma(\bm{\dn} s) 
    \right\}
    \,,
    \label{eq:smed}
\end{align}
where, for convenience, we define the complement of the decoherence parameter as $\mathcal{S}_\text{med}(t)$, which can be interpreted as the survival probability of the antenna's color configuration prior to gluon emission. 

The explicit evaluation of these quantities becomes technically challenging for realistic scattering rates $V(\bm{q}^2)$, owing to the presence of the path integral in eq.~\eqref{eq:kernel}. Although direct numerical implementations exist for single emitters~\cite{Feal:2018sml}, the standard antenna approaches yield analytical expressions by relying on additional assumptions about the parton-medium scatterings. This is typically achieved  either by restricting the resummation to soft scatterings (the harmonic oscillator approximation) or by truncating the multiple-scattering expansion to a finite number of hard scatterings (the opacity expansion).

The  multiple soft approximation corresponds to retaining the leading logarithmic contribution to the dipole cross-section~\eqref{eq:dipole},
\begin{align}
    n(s) \sigma(\bm{r}) \sim 
    \dfrac{\hat{q}(s)}{2} \bm{r}^2 + \mathcal{O}(\bm{r}^2\ln\bm{r}^2)
    \,,
\end{align}
which is expected to hold for small dipole sizes compared to the medium scale, $\bm{r}^2 \ll 1/\mu^2$. Here, $\hat{q}(s)$ is a transport coefficient characterizing the average squared transverse momentum acquired by the partons per unit path length in the medium. Within this approximation~\cite{Baier:1996kr,Zakharov:1996fv}, the momentum broadening reduces to a Gaussian distribution, while the emission kernel becomes equivalent to the propagator of a quantum harmonic oscillator with a complex frequency. For a static medium, the harmonic oscillator kernel admits a closed form expression~\footnote{Even for expanding isotropic media, the static solution remains of practical use, as scaling relations can often be established for phenomenological applications~\cite{Baier:1996kr,Salgado:2002cd,Andres:2023jao}.}.This limit is therefore referred to as Harmonic oscillator (HO) or Gaussian approximation, and is typically employed for opaque media where large transverse-size dipoles are expected to be strongly suppressed.

Another analytical scenario corresponds to an incoherent sum of a finite number of hard scatterings. Formally, this description~\cite{Gyulassy:2000er,Bethe:1953va,Wiedemann:2000za} is obtained by expansing the exponentials in~\eqref{eq:kernel}  in powers of $[n(t) \sigma(\bm{r})]^N$, and truncating the series at fixed order.  This is known as  the opacity expansion, with the $N=1$ term corresponding to the Gyulassy-Levai-Vitev (GLV) approximation~\cite{Gyulassy:2000er}.
In this context, $n(t)\sigma(\bm{r}) \sim 1/\lambda_\text{mfp}$ is the inverse mean free path of the projectile through the medium, and its time integral reduces, for a  static medium, to the opacity $L/\lambda_\text{mfp}$, where $L$ is the medium length. This approximation is therefore most appropriate for dilute media, where the series is expected to converge rapidly. 

The differences between the Gaussian approximation and opacity expansion have a direct impact on computed observables~\cite{Armesto:2011ht}. In particular, spectra obtained in the former tend to be systematically softer than those obtained in the latter, due to (i) the resummation of multiple soft scatterings and the associated interference effects, and (ii) the suppression of hard power-law tails in the HO scattering rate, which are retained in the opacity expansion. Reconciling these differences has been important for phenomenological applications, including the extraction of QGP transport coefficients and  the interpretation of their centrality and energy dependence \mbox{\cite{JET:2013cls,Andres:2016iys,Feal:2019xfl, Andres:2022bql,Apolinario:2022vzg}}.

A hybrid scheme, referred to as the Improved Opacity Expansion (IOE). was introduced in \cite{Mehtar-Tani:2019tvy}, which expands the emission kernel $\mathcal{K}$ around its harmonic oscillator solution, effectively resumming an arbitrary number of scatterings with transverse momenta below some matching scale, while treating hard scatterings above this scale order by order in opacity. The resulting framework extends the range of validity of the opacity expansion towards softer gluon emissions, at the cost of introducing a dependence on an intermediate matching scale.  For recent efforts in computing the antenna radiation spectrum within the IOE, see~\mbox{\cite{Kuzmin:2025fyu}}.

In this manuscript, we generalize the framework developed in~\cite{Andres:2020vxs} for a single color emitter to the case of a $q\bar{q}$ antenna. Within this approach, the soft-gluon emission spectrum can be obtained without further assumptions on the form of the parton–medium scattering rate, effectively resumming an arbitrary number of scatterings without relying on either the Gaussian or opacity approximations. In the next section, we show how the antenna spectrum in eq.~\eqref{eq:J-term} can be rewritten into a form suitable for numerical computation within this framework.

\section{Setting up the numerical evaluation}
\label{sec:diff-eqs}

The object of interest in this work is the medium-induced part of the antenna radiation spectrum $\mathcal{J}$. We will follow the strategy in~\cite{Andres:2020vxs}, where the momentum broadening factor $\mathcal P$ and the emission kernel $\mathcal K$ are treated as propagators satisfying Dyson-type integral equations. Substituting these equations into eq.~\eqref{eq:J-term} allows the vacuum contribution to be identified and subtracted analytically. At the same time, the integration over $\bar t$ can be traded for an integration over the time of the last medium scattering, which is bounded by the medium length. This reorganizes the interference term $\mathcal{J}$ into a form that can be evaluated directly in transverse momentum space through initial-value problems. This procedure is outlined below.

\subsection{Reorganisation of the spectrum}

In transverse momentum space, the evolution equations for the objects in eqs.~\eqref{eq:broadening}, \eqref{eq:kernel}, \eqref{eq:smed} are, respectively~\cite{Andres:2020vxs}:
\begin{align}
    \mathcal{P}(\bm{\kappa}, \tau; \bm{\ell}, \bar{t})
    &=
    (2\pi)^2\delta^2(\bm{\kappa} - \bm{\ell})
    -
    \int^{\tau}_{\bar{t}} \diff s \, n(s)
    \int_{\bm{v}}
    \mathcal{P}(\bm{\kappa}, \tau; \bm{v}, s)
    \frac{\sigma(\bm{v}-\bm{\ell})}{2}
    \,,
    \label{eq:broadening-evol}
    \\
    \begin{split}
        \mathcal{K}(\bm{q}, \bar{t}; \bm{u}, t)
        &=
        (2\pi)^2\delta^2(\bm{q} - \bm{u})
        e^{-i \tfrac{\bm{q}^2}{2\omega}(\bar{t} - t)}
        \\
        &\qquad\qquad\qquad\qquad -
        \int^{\bar{t}}_{t} \diff s \, n(s)
        \int_{\bm{v}}
        e^{-i \tfrac{\bm{q}^2}{2\omega}(\bar{t} - s)}
        \frac{\sigma(\bm{q}-\bm{v})}{2}
        \mathcal{K}(\bm{v}, s; \bm{u}, t)
        \,,
    \end{split}
    \label{eq:kernel-evol}
    \\
    \mathcal{S}_\text{med}(t) &= 
    1-
    \int^{t}_{0} \diff s\, n(s)
    \dfrac{\sigma(\bm{\dn}\,s)}{2}
    \mathcal{S}_\text{med}(s)
    \,,
    \label{eq:smed-evol}
\end{align}
where we have expressed the dipole cross-section $\sigma$ in momentum space,
\begin{align}
    \sigma(\bm{q}-\bm{\ell}) 
    = \int_{\bm{r}} e^{-i \bm{r}\cdot(\bm{q}-\bm{\ell})} \sigma(\bm{r})
    = -V(\bm{q}-\bm{\ell})
    +
    (2\pi)^2
    \delta(\bm{q}-\bm{\ell}) \int_{\bm{p}} V(\bm{p})
    \,.
\label{eq:sigma-V-momentum-space}
\end{align}
Applying these evolution equations to the interference term $\mathcal{J}$  allows us to perform the integration over  $\bar{t}$ in eq.~\eqref{eq:J-term}. A key step in this manipulation is the rearrangement of nested time integrations,
\begin{align}
\int^{\infty}_{t}\diff s_1 \,\int^{\infty}_{s_1}\diff s_2
\rightarrow
\int^{\infty}_{t}\diff s_2\int^{s_2}_{t}\diff s_1
\,,
\label{eq:rearrange}
\end{align}
together with the use of the momentum-space representation of the dipole cross-section.

Replacing eqs.~\eqref{eq:broadening-evol}--\eqref{eq:smed-evol}
into $\mathcal{J}$ generates several contributions, among which one can identify the vacuum gluon radiation out of a $q\bar{q}$ antenna:
\begin{align}
    \mathcal{J}^\text{vac}(\bm{\kappa},\bm{\bar{\kappa}})
    =
    4\omega^2
    \dfrac{\bm{\kappa}\cdot\bm{\bar{\kappa}}}{\bm{\kappa}^2\bm{\bar{\kappa}}^2}
    \,.
\end{align}

The remaining terms can then be re-written under the form
\begin{equation}
\begin{split}
\mathcal{J}^{\text{med}}(\bm{\kappa},\bm{\kappabar}) \equiv &
    \mathcal{J}(\bm{\kappa},\bm{\kappabar}) - \mathcal{J}^\text{vac}(\bm{\kappa},\bm{\kappabar}) 
    = \\
    = & \text{A}(\bm{\kappa}, \bm{\kappabar}) + \text{B}(\bm{\kappa}, \bm{\kappabar}) 
    + \frac{1}{2} \text{C}(\bm{\kappa}, \bm{\kappabar}) 
    + \frac{1}{2} \text{C}(\bm{\kappabar}, \bm{\kappa}) \,,
    \label{eq:J-results}
\end{split}
\end{equation}
upon identifying the individual contributions as follows:
\begin{align}
    \begin{split}        
        &\text{A}(\bm{\kappa}, \bm{\kappabar}) = 
        -\text{Re}
        \int^L_0 \diff s \dfrac{n(s)}{2} 
        \mathcal{S}_\text{med}(s)
        \int_{\bm{\ell}\bm{q}}
        \mathcal{P}(\bm{\kappa} - \bm{\ell}; L, s) \sigma(\bm{\ell}-\bm{q})
        \\ 
        &\hspace{.50\textwidth}
        \times e^{i s \tfrac{ \bm{q}^2 - \bm{\bar{q}}^2 }{2\omega}}
        \left[
        \mathcal{J}^{\text{vac}}(\bm{\ell},\bm{\bar{\ell}})
        -
        \mathcal{J}^{\text{vac}}(\bm{q},\bm{\bar{q}})
        \right]
        ,
    \end{split}\label{eq:A-term}
    \\
    &\text{B}(\bm{\kappa}, \bm{\kappabar}) = 
    \int^L_0 \diff s \dfrac{n(s)}{2} 
    \int_{\bm{\ell}\bm{q}}
    \mathcal{P}(\bm{\kappa} - \bm{\ell}; L, s) \sigma(\bm{\ell}-\bm{q})
    \left[
    \mathcal{J}^\text{vac}(\bm{\ell},\bm{\bar{\ell}})
    -
    \mathcal{J}^\text{vac}(\bm{q},\bm{\bar{q}})
    \right]
    ,\label{eq:B-term}
    \\
    \begin{split}
        &\text{C}(\bm{\kappa}, \bm{\kappabar}) =
        \text{Re } 
        2\omega i
        \int^L_0 \diff s 
        \,n(s)
        \int_{\bm{\ell}\bm{q}\bm{u}} 
        \int^{s}_{0} \diff t \,
        \mathcal{S}_\text{med}(t) \,
        e^{-i t \tfrac{ \bm{u}^2 - \bm{\bar{u}}^2 }{2\omega}} 
        \left(
        \dfrac{\bm{\ell}}{\bm{\ell}^2}
        -
        \dfrac{\bm{q}}{\bm{q}^2}
        \right)\cdot\bm{\bar{u}}
        \\
        &\hspace{0.45\textwidth}
        \times
        \mathcal{P}(\bm{\kappa} - \bm{\ell}; L, s) 
        \sigma(\bm{\ell}-\bm{q})
        \mathcal{K}(\bm{q},s;\bm{u},t)  
        \,.
    \end{split}
    \label{eq:C-term}
\end{align}

The resulting structure admits a natural interpretation. The $\text{A}$ and $\text{B}$ terms do not contain a medium-modified emission kernel, instead, they retain the vacuum antenna interference structure encoded in $\mathcal{J}^\text{vac}$, while all medium dependence enters only through broadening and decoherence effects.  In contrast, the $\text{C}$ terms encodes the medium-modified emission kernel.

Note that the dependence on $\bm{\bar{\kappa}}$ in eqs.~\eqref{eq:A-term}–\eqref{eq:C-term} arises from the definition of the overlined variables in~\eqref{eq:antenna_angle}. Since the $\text{A}$ and $\text{B}$ terms do not depend on the kernel $\mathcal K$, the replacement $\bm{\kappa} \to \bm{\bar{\kappa}}$ can be undone by a shift of integration variables:
$\bm{\ell + \omega\delta n} \to \bm{\ell}$, and $\bm{q + \omega\delta n} \to \bm{q}$. As a consequence, only the C-terms cannot be symmetrized under $\bm{\kappa}\leftrightarrow\bm{\kappabar}$, since the medium-modified kernel $\mathcal{K}(\bm{q},s;\bm{u},t)$  lacks translational invariance and therefore cannot be expressed solely as a function of $\bm{q-u}$.

\subsection{Decomposition of the antenna spectrum}
\label{sec:decomposition}

Having reorganised the medium-induced interference term into the contributions A, B, and C-terms in eq.~\eqref{eq:J-results}, we can now construct the medium-induced antenna differential spectrum. To this end, we follow the common procedure of splitting the spectrum into `quark-assigned' and `antiquark-assigned' contributions,
\begin{align}
    \omega\dfrac{\diff^3 I}{\diff\omega \diff^2(\bm{k}/\omega)} 
    &=
    \omega\dfrac{\diff^3 I_{q}}{\diff\omega \diff^2\bm{\nk}} 
    +
    \omega\dfrac{\diff^3 I_{\bar{q}}}{\diff\omega \diff^2\bm{\nkbar}}
    \,,
\end{align}
where the quark (antiquark) contribution $I_q$ ($I_{\bar{q}}$) is expressed as differential in the kinematic variable $\bm{\nk} \equiv \bm{\kappa}/\omega$ ($\bm{\nkbar} \equiv \bm{\kappabar}/\omega$), introduced in section~\ref{sec:qcd-antenna}.

This decomposition requires requires splitting the interference term $\mathcal{J}^\text{med}$ into two contributions,
\begin{equation}
    \mathcal{J}^\text{med} 
    = \dfrac{\mathcal{J}^\text{med}_q + \mathcal{J}^\text{med}_{\bar{q}} }{2}\ ,
\end{equation}
so that
\begin{equation}
    \omega\dfrac{\diff^3 I_{q}}{\diff\omega \diff^2\bvec{\nk}} = 
    \dfrac{\alpha_s C_F}{(2\pi)^2}
    \left[
    \mathcal{R}^\text{med}_{\,}(\bm{\kappa}) 
    - 
    \mathcal{J}^\text{med}_q(\bm{\kappa}, \bm{\kappabar})
    \right]
    \,.
\end{equation}
While the partition of the interference term is not unique, the decomposition in eq.~\eqref{eq:J-results} suggests the natural choice
\begin{align}
    \mathcal{J}^\text{med}_q(\bm{\kappa}, \bm{\kappabar})
    &\equiv
    \text{A}(\bm{\kappa}, \bm{\kappabar})
    + \text{B}(\bm{\kappa}, \bm{\kappabar})
    + \text{C}(\bm{\kappa}, \bm{\kappabar})
    \,,
    \label{eq:J-decomposition}
    \\
    \mathcal{J}^\text{med}_{\bar{q}}(\bm{\kappa}, \bm{\kappabar})
    &\equiv
    \mathcal{J}^\text{med}_q(\bm{\kappabar}, \bm{\kappa})
    \,.
\end{align}
When $\bm{\kappa}=\bm{\bar\kappa}$, corresponding to $\bm{\dn}=\bm{0}$, one has $\mathcal{S}_{\rm med}(s)=1$, and thus combination $\text{A}+\text{B}=0$. In this limit, we then have
\begin{equation}
\mathcal{R}^{\rm med}(\bm{\kappa}) 
= \text{C}(\bm{\kappa},\bm{\kappa}) \,.
\end{equation}
This choice therefore cleanly separates the independent-emission contribution from genuinely medium-induced interference effects, allowing the quark-assigned spectrum to be rewritten as
\begin{equation}
    \omega\dfrac{\diff^3 I_{q}}{\diff\omega \diff^2\bvec{\nk}} 
    \equiv \omega\dfrac{\diff^3 I^\text{Ind}}{\diff\omega \diff^2\bvec{\nk}} 
    - \omega\dfrac{\diff^3 I^\text{C}}{\diff\omega \diff^2\bvec{\nk}} 
    - \omega\dfrac{\diff^3 I^\text{AB}}{\diff\omega \diff^2\bvec{\nk}}
    \,,
    \label{eq:spectrum}
\end{equation}
where
\begin{align}
    \omega\dfrac{\diff^3 I^\text{AB}}{\diff\omega \diff^2\bvec{\nk}}
    &=
    \dfrac{\alpha_s C_F}{(2\pi)^2}
    \left[
    \text{A}(\bm{\kappa}, \bm{\kappabar}) 
    + \text{B}(\bm{\kappa},\bm{\kappabar})
    \right]
    \,,\\
    \omega\dfrac{\diff^3 I^\text{C}}{\diff\omega \diff^2\bvec{\nk}}
    &=
    \dfrac{\alpha_s C_F}{(2\pi)^2}
    \text{C}(\bm{\kappa}, \bm{\kappabar})
    \,,\\
    \omega\dfrac{\diff^3 I^\text{Ind}}{\diff\omega \diff^2\bvec{\nk}}
    &=
    \dfrac{\alpha_s C_F}{(2\pi)^2}
    \text{C}(\bm{\kappa}, \bm{\kappa}) 
    \, .
\end{align}
The quantity $I^\text{Ind}$ corresponds to the independent-emission limit of the antenna spectrum. 
This decomposition thus separates contributions including medium-induced modifications of the emission kernel ($I^\text{C}$) from those involving only the momentum broadening and decoherence factors ($I^\text{AB}$). 

Finally, we can obtain the azimuthally-integrated distribution:
\begin{align}
    \omega\dfrac{\diff^2 I}{\diff\omega \diff\nk}
    =
    \nk
    \int^{2\pi}_{0} \diff\varphi^{}_\kappa
    \,
    \omega\dfrac{\diff^3 I}{\diff\omega \diff^2\bvec{\nk}}
    \,,
\label{eq:azimuthal-sum-spectrum}
\end{align}
where $\varphi^{}_\kappa$ is taken as the angle between $\bm{\kappa}$ (or $\bm{\nk}$) and $\bm{\dn}$\footnote{In general, we will use $\varphi_q$ for the angle between any momentum variable $\bm{q}$ and the $\bm{\dn}$ direction.}.

It is worth noting that, while the decomposition in eq.~\eqref{eq:J-decomposition} is not unique\footnote{In fact, it differs from the usual convention, where the `quark-assigned' contribution is taken proportional to $\mathcal{R}(\bm{\kappa}^2) - \mathcal{J(\bm{\kappa},\bm{\kappabar})}$, as in the vacuum case (see, e.g.~\cite{Mehtar-Tani:2012mfa,Kuzmin:2025fyu}). Although both prescriptions reproduce the fully differential antenna spectrum, they lead to different azimuthally integrated quark and antiquark-assigned  contributions, which can complicate quantitative comparisons.},
it can be motivated in two ways. First,  contributions in which the medium-modified kernel $\mathcal{K}$ is evaluated with transverse momenta relative to the quark are assigned to the quark contribution (and analogously for the antiquark). Second, terms proportional to the vacuum interference term $\mathcal{J}^\text{vac}$, namely the $\text{A}$ and $\text{B}$ terms, are shared equally between $I_q$ and $I_{\bar{q}}$, ensuring that `anti-angular ordered' radiation is present in both contributions.

\subsection{Initial value problems}

In this section, we present the numerical methods used to evaluate each of the contributions to the medium-induced spectrum off an antenna. To this end, we start by recasting the evolution equations for the emission kernel and momentum broadening into an initial value problem.

For momentum broadening, one obtains
\begin{align}
    \partial_\tau \mathcal{P}(\bm{\kappa}-\bm{\ell};\tau,s) 
    &=
    -\dfrac{1}{2}n(\tau)
    \int_{\bm{v}}
    \sigma(\bm{\kappa}-\bm{v})
    \mathcal{P}(\bm{v}-\bm{\ell};\tau,s)
    \,,
    \label{eq:broadening-evol-diff}
    \\
    \mathcal{P}(\bm{\kappa}-\bm{\ell};\tau=s,s)
    &= (2\pi)^2 \delta(\bm{\kappa}-\bm{\ell})
    \,,
\end{align}
and for the emission kernel,
\begin{align}
    \partial_t 
    \mathcal{K}(\bm{q},s;\bm{u},t) 
    &=
    i \dfrac{\bm{u}^2}{2\omega}
    \mathcal{K}(\bm{q},s;\bm{u},t) 
    +\dfrac{1}{2}n(t)
    \int_{\bm{v}}
    \mathcal{K}(\bm{q},s;\bm{v},t) 
    \sigma(\bm{v}-\bm{u})
    \,,
    \label{eq:kernel-evol-diff}
    \\
    \mathcal{K}(\bm{q},s;\bm{u},t=s) 
    &= 
    (2\pi)^2 \delta(\bm{q}-\bm{u})
    \,.
\end{align}
For convenience, we introduce specific combinations of the medium and antenna parameters. In particular, we replace $n_0$, $L$, $\mu$, and $\dn$ by 
\begin{align}
    n_0 L \,,\quad 
    \omegaC = \dfrac{\mu^2 L}{2} \,,\quad 
    \thetaC = \dfrac{2}{\mu L} = \dfrac{\mu}{\omegaC}\,,\quad 
    \RC =\dfrac{\dn}{\thetaC} \,, 
\label{eq:dimensionless-scales}
\end{align}
where $\omegaC$ sets the characteristic gluon energy scale and $\thetaC$ the critical antenna angle. We also express the gluon kinematics as a function of the rescaled (dimensionless) variables
\begin{equation}
x=\omega/\omegaC\,, \quad \rk = \kappa/(\omega\dn) = \nk/\dn \,.
\end{equation}

For the azimuthal integration over transverse momenta, we exploit the rotational invariance of the collision rate, $V(\bm{\kappa}-\bm{q}) =V(|\bm{\kappa}-\bm{q}|) = V(\kappa,q,\cos\varphi_{\kappa q})$, where $\varphi_{\kappa q}$ denotes the relative angle between $\bm{\kappa}$ and $\bm{q}$. This allows us to the collision rate in angular harmonics of $\varphi_{\kappa q}$
\begin{align}
    M_n(\kappa,q;\mu) 
    &=
    \int^{2\pi}_{0} \dfrac{\diff\varphi_{\kappa q}}{2\pi}
    \dfrac{V(\kappa,q,\cos\varphi_{\kappa q};\mu)}{4\pi}
    \cos(n\varphi_{\kappa q})
    \,,
\label{eq:harmonics-def}
    \\
    \dfrac{V(\kappa,q,\cos\varphi_{\kappa q};\mu)}{4\pi}
    &=
    M_0(\kappa, q;\mu)
    +
    2 \sum_{n=1}^{\infty} M_n(\kappa, q;\mu) \cos(n\varphi_{\kappa q})
    \,.
\end{align}
where $\mu$ is the Debye screening mass parameter, set by the thermal scale of the medium. These harmonics admit closed analytical expressions for both the Yukawa and Hard Thermal Loop scattering rates, which are given in appendix~\ref{app:averages-details}. We note that in the notation of~\cite{Andres:2020vxs}, they correspond to $M_0 \equiv V_1/4\pi$ and $M_1 \equiv V_2 / 4\pi$. 

\subsubsection{The decoherence parameter}
\label{sec:smed}

The decoherence parameter $\Delta_\text{med} = 1-\Smed$, introduced in eq.~\eqref{eq:smed}, can be evaluated directly from the dipole cross-section $\sigma(\bm{r})$, as:
\begin{equation}
\mathcal{S}_{\text{med}}(t)
=
\begin{cases}
\exp\!\left[-\displaystyle\int_0^t \mathrm{d}s \, \frac{n(s)}{2}\, \sigma_{\text{Yuk}}(\bm{\dn}s)\right],
& \text{Yukawa model,} \\[0.8em]
\exp\!\left[-\displaystyle\int_0^t \mathrm{d}s \, \frac{1}{2}\, \alpha_s N_c T(s)\, \sigma_{\text{HTL}}(\bm{\dn}s)\right],
& \text{HTL model.}
\end{cases}
\label{eq:smed-yuk-htl}
\end{equation}
In each case, the medium dependence is separated into a density profile (either $n$ or $T$) and the dipole cross-section, which is computed from the corresponding scattering rate $V(\bm{q}^2)$ via eq.~\eqref{eq:dipole}. Explicitly, the Yukawa and HTL interaction rates are respectively given by
\begin{align}
    V_\text{Yuk}(\bm{q}^2, \mu^2) &=
    \dfrac{8\pi \mu^2}{(\bm{q}^2 + \mu^2)^2}
    \,,\\
    V_\text{HTL}(\bm{q}^2, \md^2) &= 
    \dfrac{8\pi \md^2}{(\bm{q}^2 + \md^2) \bm{q}^2}
    \,,
\end{align}
where $\mu$ and $m_D$ set the characteristic transverse-momentum scales of the interaction. Although both models share the same ultraviolet behaviour, $V(\bm{q}^2)\propto \bm{q}^{-4}$ for $\bm{q}^2 \gg \mu^2, m_D^2$, reflecting the Coulomb tail of the interaction, they differ in the infrared region, which leads to different momentum-broadening and decoherence patterns.

To compare the Yukawa and HTL descriptions on equal footing, one must specify a mapping between their respective opacity parameters ($n_0$, $T$) and screening masses ($\mu^2$, $\md^2$). A first condition is obtained from matching the large transverse-momentum behavior of the two models, which yields $\alpha_s N_c T \md^2 = n_0 \mu^2$. This condition ensures that  the exponents of~\eqref{eq:smed-yuk-htl} coincide in the ultraviolet regime. A second relation is fixed by following the prescription of~{\cite{Barata:2020sav}}, where it is shown that the Yuakawa and HTL dipole cross-sections can be matched at leading logarithmic accuracy for small dipole sizes by also imposing the condition $\md^2 = e \mu^2$.  Together, these relations define a one-to-one correspondence between the Yukawa and HTL medium parameters serving,  allowing us to express the dimensionless HTL parameters using the Yukawa ones as
\begin{align}
    \alpha_s N_c TL = \dfrac{n_0 L}{e}
    \,,\quad
    \omegaC^\text{H} = \dfrac{\md^2 L}{2} = e \omegaC
    \,,\quad
    \thetaC^\text{H} = \dfrac{2}{\md L} = \dfrac{\thetaC}{\sqrt{e}}
    \,,\quad
    \RC^\text{H} = \dfrac{\dn}{\thetaC^\text{H}} = \sqrt{e} \RC
    \,.
\label{eq:HTL-Yukawa-matching}
\end{align}
For the remainder of this study, all results will be expressed in terms of the Yukawa quantities: $n_0L, \omegaC, \thetaC, \RC$.

For completeness, we present the dipole cross-sections entering $\Smed$ for the Yukawa and Hard Thermal Loop collision rates:
\begin{align}
	\dfrac{1}{2}\sigma_\text{Yuk}(r)
	&=
	1 - \mu r \, K_1(\mu r)
	\,,
	\\
	\dfrac{1}{2}\sigma_\text{HTL}(r)
	&=
	2 \left[
	K_0(\md r) + \ln \dfrac{\md r}{2} + \gamma_\text{E}
	\right]
	\,,
\end{align}
where $K_\nu$ are the modified Bessel functions of the second kind with order $\nu$ and $\gamma_\text{E}$ is the Euler-Mascheroni constant.

For a medium with length $L$ and constant density $n(t) = n_0 \Theta(L-t)$, the exponent in  $\Smed(t)$, requires the following integrals:
\begin{flalign}
	\int^z_0 
	\hspace{-2pt}
	\diff u 
	\left[ 1 - u \, K_1(u) \right]
	&=
	z \left[
	1 + K_0(z)
	- \dfrac{\pi}{2} 
	\left(
	L_{-1}(x) K_0(z)
	+
	L_{0}(x) K_1(z)
	\right)
	\right]
	,
	\\
	\hspace{-6pt}
	\int^z_0  
	\hspace{-2pt}
	\diff u 
	\left[ 
	K_0(u) + \ln \dfrac{u}{2} + \gamma_\text{E}
	\right]
	\hspace{-2pt}
	&=
	z \left[
	\gamma_\text{E} - 1
	+
	\ln \dfrac{z}{2} 
	+ \dfrac{\pi}{2} 
	\left(
	L_{-1}(z) K_0(z)
	+
	L_{0}(z) K_1(z)
	\right)
	\right]
	\hspace{-2pt}
	.
	\hspace{-3pt}
\end{flalign}
\noindent where $L_\nu$ the modified Struve functions\footnote{%
The relevant property for these integrals is $\int K_0(u) \diff u = \tfrac{\pi}{2}(L_{-1}(u) K_{0}(u) + L_{0}(u) K_{1}(u))$.
} of order $\nu$.

\subsubsection{The A and B terms}
Since our final goal is to obtain the azimuthally integrated spectrum, it is convenient to introduce azimuthally averaged auxiliary functions. We first consider the contribution to the spectrum coming from the A and B terms. In the original kinematic variables $(\omega, \bm{\nk}/\dn)$, it reads
\begin{align}
    -\omega \dfrac{\diff^3 I^\text{AB}}{\diff\omega\diff^2(\bm{\nk}/\dn)}
    =
    \dfrac{\alpha_s C_F}{(2\pi)^2}
    \int^{L}_{0} \diff s \,n(s) \,
    \psi^{\,}_\text{AB}(
    \bm{\kappa}=\omega\bm{\nk},
    \omega;
    \tau=L,s)
    \,,
\end{align}
where we have introduced the function $\psi^{\,}_{\rm AB}$ defined by 
\begin{equation}
    \psiAB(\bm{\kappa},\omega;\tau,s) 
    =
    \int_{\bm{\ell}} 
    \mathcal{P}(\bm{\kappa} - \bm{\ell}; \tau, s)
    \left[
    \Psi(\bm{\ell},\omega;s=0)
    -
    \Smed(s)
    \Psi(\bm{\ell},\omega;s)
    \right]
    \,,
    \label{eq:psiAB-def}
\end{equation}
with the  auxiliary function $\Psi(\bm{\ell},\omega;s)$   given by
\begin{equation}
    \Psi(\bm{\ell},\omega;s)
    =
    \dn^2
    \int_{\bm{q}}
    \dfrac{V(\bm{\ell}-\bm{q})}{2}
    \left[
    \mathcal{J}^{\text{vac}}(\bm{\ell},\bm{\bar{\ell}})
    -
    \mathcal{J}^{\text{vac}}(\bm{q},\bm{\bar{q}})
    \right]
    \,\text{Re }
    e^{i s \tfrac{ \bm{q}^2 - \bm{\bar{q}}^2 }{2\omega}}
    \,.
\end{equation}
We now perform the azimuthal integration which yields, as a function of the variables $\omega$, and $\rk$,
\begin{align}
    -\omega \dfrac{\diff^2 I^\text{AB}}{\diff\omega\diff\rk}
    =
    \dfrac{\alpha_s C_F}{2\pi}
    \rk
    \int^{L}_{0} \diff s \,n(s) \,
    \tilde{\psi}^{\,}_\text{AB}(
    \kappa=\omega\dn\,\rk,
    \omega;
    \tau=L,s)
    \,,
\end{align}
where $\psiABavg$ represents the azimuthally averaged version of $\psiAB$
\begin{align}
    \psiABavg(\kappa,\omega;\tau,s)
    &=
    \int^{2\pi}_{0} \dfrac{\diff\varphi_\kappa}{2\pi}
    \psiAB(\bm{\kappa},\omega;\tau,s)
    \,,
    \label{eq:azimuthal-average-psiAB}
    \\
    \tilde{\Psi}(\kappa,\omega;s) 
    &=
    \int^{2\pi}_{0} \dfrac{\diff\varphi_\kappa}{2\pi}
    \Psi(\bm{\kappa},\omega;s)
    \,.
    \label{eq:azimuthal-average-Psi}
\end{align}
Proceeding now to rescale all variables into dimensionless quantities, we define new auxiliary functions such that:
\begin{align}
    \gAB(\rk,x;\bar{\tau},\bar{s})
    &=
    \psiABavg(\rk \omega\dn,x \omegaC;
    \bar{\tau} L,\bar{s} L 
    )
    \,,\\
    \GAB
    (\rk,x;\bar{s})
    &=
    \tilde{\Psi}(\rk \omega\dn,x \omegaC;\bar{s} L )
    \,,
\end{align}
where $x=\omega/\omegaC$ is the rescaled energy, while $\bar{s} = s/L$ and $\bar{\tau}=\tau/L$ are rescaled time coordinates.

Using the broadening evolution equation~\eqref{eq:broadening-evol-diff}, performing the azimuthal average, and expressing all quantities in terms of the dimensionless variables introduced above, one obtains the evolution equation: 
\begin{equation}
    \partial_{\bar{\tau}}
    \gAB(\rk,x;\bar{\tau},\bar{s})
    =
    -n(\bar{\tau})L
    \int^\infty_0 \diff u \, u
 \,    M_0\Big(\rk,u;\frac{1}{\RC x}\Big)
    \big[
    \gAB(\rk,x;\bar{\tau},\bar{s})
    -
    \gAB(u,x;\bar{\tau},\bar{s})
    \big]
    \, ,
\end{equation}
\noindent where we made use of the following scaling property of $M_n$: 
\begin{equation}
    M_n(\kappa,q;\mu) q \, \diff q =
    M_n \left( \frac{\kappa}{\omega \dn}, \frac{q}{\omega \dn}; \frac{\mu}{\omega \dn} \right) \frac{q}{\omega \dn} \diff\left( \frac{q}{\omega \dn} \right)
    \,.
    \label{eq:harmonic-rescale-property}
\end{equation}

What remains is to determine the initial condition for $\gAB(\rk,x;\bar{\tau},\bar{s})$ which can be set at $\bar{\tau}=\bar{s}$, in eq.~\eqref{eq:psiAB-def}. This results in:
\begin{align}
    \gAB(\rk,x;\bar{\tau}=\bar{s})
    &=
    \GAB(\rk,x;\bar{s}=0)-
    \Smed(\bar{s}L) 
    \GAB(\rk,x;\bar{s})
    \,,
\end{align}
such that the remaining task is the analytical evaluation of
$\GAB(\rk,x;\bar s)$.
After performing the angular integrations analytically, one obtains (details can be found in appendix~\ref{app:averages-details}):
\begin{align}
\begin{split}
     \GAB(\rk,x;\bar{s})
     =&
     \cos(\RC^2 x \bar{s})
     \times
     \bigg[
     \mathcal{I}_{0}(\RC x, \RC \bar{s}, \rk)
     +
     2
     \sum^\infty_{j=1} (-1)^j
     \mathcal{I}_{2j}(\RC x, \RC \bar{s}, \rk)
     \bigg]
     \\
     +&
     \sin(\RC^2 x \bar{s})
     \times
     2
     \sum^\infty_{j=1} (-1)^j
     \mathcal{I}_{2j-1}(\RC x, \RC \bar{s}, \rk)
     \,,
\end{split}
\label{eq:H-init-condition}
\end{align}
\noindent where the integrals $\mathcal{I}_n$ are defined in terms of the scattering rate harmonics $M_n$ defined in eq.~\eqref{eq:harmonics-def}, and the Bessel functions of the first kind $J_n$:
\begin{align}
\begin{split}
    \mathcal{I}_n(\RC x, \RC \bar{s}, \rk) 
   = & \,
    \dfrac{4}{r^2_\kappa}
    \int \dfrac{\diff u}{u}
    J_n(2\RC^2 x \bar{s}\,u)
    \times
    \\
    \times
    &\left[
    u^2 D_n(\rk) M_n\left(r_\kappa, u; \frac{1}{\RC x}\right)
    - 
    \rk^2 D_n(u) M_0\left(r_\kappa, u; \frac{1}{\RC x}\right)
    \right]
    \,,
\label{eq:I-integrals-def}
\end{split}\\
    D_{0}(y) 
    = & \, \Theta(y-1) \,,
    \\
    D_{n}(y) 
   = & \, 
    \dfrac{
    \Theta(y-1) y^{-n} - \Theta(1-y) y^{n} 
    }{2}
    \,, \quad n \geq 1
    \,.
\end{align}

Finally, putting all the ingredients together, the spectrum contribution in terms of the rescaled variables reads:
\begin{align}
    -x \dfrac{\diff^2 I^\text{AB}}{\diff x\diff (\theta/\dn)}
    =
    \dfrac{\alpha_s C_F}{2\pi}
    \frac{\theta}{\dn}
    \int^{1}_{0} \diff \bar{s} \,n(\bar{s})L \,
    \gAB\left( \frac{\theta}{\dn},x;\bar{\tau}=1,\bar{s} \right)
    \,,
    \label{eq:spec-AB-adimensional}
\end{align}
\noindent where $\gAB(\rk,x;1,\bar{s})$ is the solution to the initial value problem:
\begin{align}
    \partial_{\bar{\tau}}
    \gAB(\rk,x;\bar{\tau}, \bar{s})
    &=
    -n(\bar{\tau})L
    \int^\infty_0 \diff u u
    M_0\Big(\rk,u;\frac{1}{\RC x}\Big)
    \left[
    \gAB(\rk,x;\bar{\tau}, \bar{s})
    -
    \gAB(u,x;\bar{\tau}, \bar{s})
    \right]
    \,,
    \label{eq:h-evol-eq}
    \\
    \gAB(\rk,x;\bar{\tau}=\bar{s},\bar{s})
    &=
    \GAB(\rk,x;\bar{s}=0)-
    \Smed(\bar{s}L) 
    \GAB(\rk,x;\bar{s})
    \,.
    \label{eq:h-init-cond}
\end{align}

The procedure to obtain the $I^\text{AB}$ spectrum contribution is thus as follows: evaluate the initial condition\footnote{
Evaluating the initial condition requires truncating the infinite series in eq.~\mbox{\eqref{eq:H-init-condition}}. In practice, four terms of each alternating sum are sufficient to ensure convergence over the entire phase-space.
} 
in eq.~\eqref{eq:h-init-cond} for all values of $\bar{s}\in[0,1]$, solve eq.~\eqref{eq:h-evol-eq} from $\bar{\tau}=\bar{s}$ to $\bar{\tau}=1$, and finally integrate the result in $\bar{s}$, according to eq.~\eqref{eq:spec-AB-adimensional}.

\subsubsection{The C term}

We now turn to the evaluation of the $\text{C}$ contribution, which contains the \st{genuinely} medium-modified emission kernel. In contrast to the $\text{A}$ and $\text{B}$ terms, this contribution depends simultaneously on momentum broadening and in-medium propagation, requiring the solution of coupled evolution equations. Following the strategy of refs.~\cite{Caron-Huot:2010qjx,Andres:2020vxs}, we introduce a set of auxiliary vector functions that reorganize the problem into a sequence of initial value problems. Namely:
\begin{align}
    \bm{\phi}(\bm{\kappa},\tau;\bm{q},s)
    &=
    \int_{\bm{\ell}}
    \mathcal{P}(\bm{\kappa}-\bm{\ell};\tau,s)
    \sigma(\bm{\ell}-\bm{q})
    \left(
    \dfrac{\bm{\ell}}{\bm{\ell}^2}
    -
    \dfrac{\bm{q}}{\bm{q}^2}
    \right)
    \,,\\
    \bm{\psi}(\bm{\kappa};s;\bm{u},t)
    &=
    \int_{\bm{q}}
    \bm{\phi}(\bm{\kappa},\tau=L;\bm{q},s)
    \mathcal{K}(\bm{q},s;\bm{u},t)
    \,,
\end{align}
The $\text{C}$ contribution is therefore expressed in terms of $\bm{\psi}$, itself constructed from $\bm{\phi}$, with both quantities determined through initial value problems. For $\bm{\phi}$ the broadening evolution in eq.~\eqref{eq:broadening-evol-diff} becomes:
\begin{align}
    \partial_\tau 
    \bm{\phi}(\bm{\kappa},\tau;\bm{q},s)
    &=
    -\dfrac{1}{2}n(\tau)
    \int_{\bm{y}}
    \sigma(\bm{\kappa}-\bm{y})
    \bm{\phi}(\bm{y},\tau;\bm{q},s)
    \,,
    \\
    \bm{\phi}(\bm{\kappa},\tau=s;\bm{q},s)
    &= 
    \sigma(\bm{\kappa}-\bm{q})
    \left(
    \dfrac{\bm{\kappa}}{\bm{\kappa}^2}
    -
    \dfrac{\bm{q}}{\bm{q}^2}
    \right)
    \,,
\end{align}
\noindent whereas for $\bm{\psi}$ the kernel evolution in eq.~\eqref{eq:kernel-evol-diff} turns into:
\begin{align}
    \partial_t 
    \bm{\psi}(\bm{\kappa};s;\bm{u},t)
    &=
    i \dfrac{\bm{u}^2}{2\omega}
    \bm{\psi}(\bm{\kappa};s;\bm{u},t)
    +\dfrac{1}{2}n(t)
    \int_{\bm{y}}
    \bm{\psi}(\bm{\kappa};s;\bm{y},t)
    \sigma(\bm{y}-\bm{u})
    \,,
    \label{eq:psi-evol}
    \\
    \bm{\psi}(\bm{\kappa};s;\bm{u},t=s)
    &= 
    \bm{\phi}(\bm{\kappa},\tau=L;\bm{u},s)
    \,.
\end{align}
Although these equations fully determine the $\text{C}$ contribution, two practical complications remain. One issue is the first term in eq.~\eqref{eq:psi-evol}, responsible for oscillatory behavior in $\bm{\psi}$. This problem was already noted in~\cite{Caron-Huot:2010qjx}, and addressed by changing to an `interaction picture' representation:
\begin{align}
    \bm{\psi}_\text{I}(\bm{\kappa};s;\bm{u},t)
    &=
    \bm{\psi}(\bm{\kappa};s;\bm{u},t)
    e^{i(s-t)\tfrac{\bm{u}^2}{2\omega}}
    \,,
    \\
    \partial_t 
    \bm{\psi}_\text{I}(\bm{\kappa};s;\bm{u},t)
    &=
    +\dfrac{1}{2}n(t)
    \int_{\bm{y}}
    \bm{\psi}_\text{I}(\bm{\kappa};s;\bm{y},t)
    \sigma(\bm{y}-\bm{u})
    e^{i(s-t)\tfrac{\bm{u}^2-\bm{y}^2}{2\omega}}
    \,,\\
    \bm{\psi}_\text{I}(\bm{\kappa};s;\bm{u},t=s)
    &= 
    \bm{\phi}(\bm{\kappa},\tau=L;\bm{u},s)
    \,,
\end{align}
allowing us to write the $\text{C}$ term as a function of $\bm{\psi}_\text{I}$:
\begin{align}
    \omega\dfrac{\diff^3 I^\text{C}}{\diff\omega \diff^2\bvec{\nk}}
    =
    \dfrac{\alpha_s C_F}{(2\pi)^2}
    \text{Re}
    \,2\omega i
    \int^L_0 \diff s 
    \,n(s)
    \int^{s}_{0} \diff t
    \mathcal{S}_\text{med}(t)
    \int_{\bm{u}} 
    e^{-i \left( 
    s \frac{\bm{u}^2}{2\omega} 
    - t \frac{\bm{\bar{u}}^2}{2\omega} \right)} 
    \bm{\psi}_\text{I}(\bm{\kappa};s;\bm{u},t)
    \cdot\bm{\bar{u}} \,.
\end{align}
Another issue is the large number of degrees of freedom in $\bm{\psi}_\text{I}$, due to the angular dependences on $\bm{\kappa}$ and $\bm{u}$. We circumvent this issue by focusing only on the azimuthally integrated antenna spectrum. Thus, we define the functions:
\begin{align}
    \int^{2\pi}_0 \dfrac{\diff\varphi_\kappa}{2\pi}
    \bm{\psi}_\text{I}(\bm{\kappa};s;\bm{u},t)
    &=
    \tilde{\psi}_\text{I}(\kappa;s;u,t) 
    \dfrac{\bm{u}}{\bm{u}^2}
    \,,
    \\
    \int^{2\pi}_0 \dfrac{\diff\varphi_\kappa}{2\pi}
    \bm{\phi}(\bm{\kappa},\tau;\bm{q},s)
    &=
    \tilde{\phi}(\kappa,\tau;q,s)
    \dfrac{\bm{q}}{\bm{q}^2}
    \,.
\end{align}
After azimuthal averaging, the remaining angular dependence can be integrated analytically, yielding
\begin{align}
\begin{split}
    \omega\dfrac{\diff^2 I^\text{C}}{\diff\omega \diff\nk}
    =
    \dfrac{\alpha_s C_F}{2\pi}
    2\omega\nk
    \,\text{Re}
    \, i
    \int^L_0 \diff s 
    \,n(s)
    \int^{s}_{0} \diff t 
    \int \dfrac{\diff u \,u}{2\pi}
    \,
    e^{-i (s-t) \tfrac{u^2}{2\omega} } 
    &
    \tilde{\psi}_\text{I}(\kappa=\omega\theta;s;u,t)
    \,\times
    \\
    \times\,\Smed(t)
    &
    \Sigma_\text{coh}\left(t\dfrac{\dn^2 \omega}{2}, t\dn u\right)
    ,
\end{split}
\label{eq:averaged-IC}
\end{align}
such that the entire effect of the interference term is captured in the functions $\Smed$ and $\Sigma_\text{coh}$, where the latter is defined as
\begin{align}
    \Sigma_\text{coh}(\lambda,\rho)=
    e^{i\lambda}
    \left[
    J_0(\rho) + 
    2i\lambda
    \dfrac{J_1(\rho)}{\rho}
    \right]
    \,.
\end{align}

The spectrum of independent radiation from the quark (or antiquark) can be recovered by evaluating eq.~\eqref{eq:averaged-IC} at $\dn=0$:
\begin{align}
    \omega\dfrac{\diff^2 I^\text{Ind}}{\diff\omega \diff\nk}
    =
    \dfrac{\alpha_s C_F}{2\pi}
    2\omega\nk
    \,\text{Re}
    \, i
    \int^L_0 \diff s 
    \,n(s)
    \int^{s}_{0} \diff t 
    \int \dfrac{\diff u \,u}{2\pi}
    e^{-i (s-t) \tfrac{u^2}{2\omega} } 
    \tilde{\psi}_\text{I}(\kappa=\omega\theta;s;u,t)
    \,,
\end{align}

We have therefore reduced this problem to computing the function $\tilde{\psi}_\text{I}$. Performing the angular integrations, we find the following initial value problem for the broadening function $\tilde{\phi}$:
\begin{align}
    \partial_\tau 
    \tilde{\phi}(\kappa,\tau;q,s) 
    &=
    -n(\tau)
    \int^\infty_0 \diff y \, y
    \, M_0(\kappa,y;\mu)
    \left[
    \tilde{\phi}(\kappa,\tau;q,s)
    -
    \tilde{\phi}(y,\tau;q,s)
    \right]
    \,,
    \\
    \tilde{\phi}(\kappa,\tau=s;q,s) 
    &= 4\pi
    \left[
    M_0(\kappa,q;\mu)
    -
    \dfrac{q}{\kappa}
    M_1(\kappa,q;\mu)
    \right]
    \,,
\end{align}
\noindent and for the kernel function $\tilde{\psi}_\text{I}$:
\begin{align}
\begin{split}
    \partial_t 
    \tilde{\psi}_{\text{I}}(\kappa;s;u,t) 
    &=
    +n(t)
    \int^\infty_0 \diff y
    \Big[
    y
    M_0(u,y;\mu)
    \tilde{\psi}_{\text{I}}(\kappa;s;u,t) 
    \\
    &\qquad\qquad\qquad\qquad
    -
    u
    M_1(u,y;\mu)
    e^{i(s-t)\tfrac{u^2-y^2}{2\omega}}
    \tilde{\psi}_{\text{I}}(\kappa;s;y,t) 
    \Big]
    \,,
\end{split}
\label{eq:kernel-evol-averaged}
    \\
    \tilde{\psi}_{\text{I}}(\kappa;s;u,t=s)
    &=
    \tilde{\phi}(\kappa,L;u,s)
    \,.
\end{align}

Finally we can re-scale the arguments of both functions and restate the problem in dimensionless variables, e.g.:
\begin{align}
    t \to tL \,,\quad
    K = \kappa/\mu \,,\quad
    \nu = u\sqrt{\dfrac{L}{2\omega}}
    = \dfrac{u/\mu}{\sqrt{x}}
    \,,
\end{align}
\noindent such that the time coordinates $t,s,\tau$ are expressed in units of the medium length $L$, the broadening arguments $\kappa,q$ in units of the screening mass $\mu$, and the kernel argument $u$ is rescaled to simplify the phase in eq.~\eqref{eq:kernel-evol-averaged}. This yields the rescaled functions $G$ and $f_x$, which correspond respectively to the rescaled broadening and kernel evolution functions:
\begin{align}
    G(K,\tau;Q,s) 
    &=
    \tilde{\phi}(K\mu,\tau L;Q\mu,s L) 
    \dfrac{\mu^2}{4\pi}
    \,,
    \\
    f_{x}(K;s;\nu,t)
    &=
    \tilde{\psi}_\text{I}    \left(K\mu;sL;\nu\sqrt{x}\mu,tL\right) 
    \dfrac{\mu^2}{4\pi}
    \,,
\end{align}
\noindent whose evolution equations and initial conditions read, for the broadening evolution:
\begin{align}
    \dfrac{\partial}{\partial\tau}
    G(K,\tau;Q,s)
    &=
    -n(\tau)L
    \int^\infty_0 \diff y \, y \,M_0(K,y;1)
    \left[
    G(K,\tau;Q,s)
    -
    G(y,\tau;Q,s)
    \right]
    \,,\\
    G(K,\tau=s;Q,s)
    &=
    M_0(K,Q;1) - \dfrac{Q}{K} M_1(K,Q;1)
    \,,
\label{eq:broadening-evol-dimensionless}
\end{align}
\noindent and for the kernel evolution:
\begin{align}
\begin{split}
    \dfrac{\partial}{\partial t}
    f_{x}(K;s;\nu,t)
    &=
    n(t)L
    \int^{\infty}_0
    \diff y
    \Big[
    y
    M_0(\nu,y;1/\sqrt{x})
    f_{x}(K;s;\nu,t) 
    \\
    &\qquad\qquad\qquad\qquad
    -
    \nu
    M_1(\nu,y;1/\sqrt{x})
    e^{i(s-t)(\nu^2-y^2)}
    f_{x}(K;s;y,t)
    \Big]
    \,,
\end{split}
    \\
    f_{x}(K;s;\nu,t=s)
    &=
    G(K,\tau=1;\nu\sqrt{x},s)
    \,.
\end{align}
In obtaining the dimensionless form of the evolution equations, we again make use of the scaling property of the scattering rate harmonics given in eq.~\eqref{eq:harmonic-rescale-property}, but now considering $(\mu \sqrt{x})^{-1}$ as pre-factor.

In terms of this rescaled function, the spectrum contributions read:
\begin{align}
\begin{split}
    x\dfrac{\diff^2 I^\text{C}}{\diff x \diff(\nk/\thetaC)}
    =
    \dfrac{4\alpha_s C_F}{\pi}
    \frac{x^2\nk}{\thetaC}
    \text{Re}
    \int^1_0 
    \hspace{-0.10cm}
    \diff s \,
    n(sL)L
    \int^{s}_{0} 
    \hspace{-0.10cm}
    \diff t 
    \int^\infty_0 \diff \nu \nu
    &
    i
    e^{-i (s-t) \nu^2 } 
    f_x\Big(x\dfrac{\nk}{\thetaC};s;\nu,t\Big)
    \\
    \times\,\Smed(tL)
    &
    \Sigma_{\text{coh}}\left(
    t\RC^2 x, 
    2t\RC\nu\sqrt{x}
    \right)
    ,
\end{split}
\label{eq:C-spec}
    \\
    x\dfrac{\diff^2 I^\text{Ind}}{\diff x \diff(\nk/\thetaC)}
    =
    \dfrac{4\alpha_s C_F}{\pi}
    \frac{x^2\nk}{\thetaC}
    \text{Re}
    \int^1_0 
    \hspace{-0.10cm}
    \diff s \,
    n(sL)L
    \int^{s}_{0} 
    \hspace{-0.10cm}
    \diff t 
    \int^\infty_0 \diff \nu \nu
    &i
    e^{-i (s-t) \nu^2 } 
    f_x\Big(x\dfrac{\nk}{\thetaC};s;\nu,t\Big)
    .
\label{eq:C0-spec}
\end{align}

\subsection{Computing the GLV spectra in a QGP \textit{brick}}
\label{sec:first-opacity}

We now present semi-analytical expressions for the antenna emission spectrum in the first-opacity limit, which will be compared to our fully resummed framework in the next section. The GLV approximation can be recovered from our results by replacing the broadening and kernel functions in eqs.~{\eqref{eq:A-term}},~{\eqref{eq:B-term}}, and~{\eqref{eq:C-term}} by their vacuum forms (see~\cite{Andres:2020vxs}). This is equivalent to evaluating the spectra in eqs.~{\eqref{eq:spec-AB-adimensional}},~{\eqref{eq:C-spec}}, and~{\eqref{eq:C0-spec}} using the initial conditions for each corresponding initial value problem.
Following this approach, we provide the expressions for the GLV spectrum assuming for a medium of length fixed length $L$ and a constant linear density of scattering centres: $n(t)=n_0\Theta (L-t)$, as this is the medium model used for the comparison between the full and GLV approaches in the next section. 
Under the single-scattering approximation some of the time integrations can be performed analytically within this brick, yielding the following expressions for the spectrum contributions
\begin{align}
    -x \dfrac{\diff^2 I^\text{AB}_\text{GLV}}{\diff x\diff (\theta/\dn)}
    &=
    \dfrac{\alpha_s C_F}{2\pi}
    \frac{\theta}{\dn}
    n_0 L
    \int^{1}_{0} \diff s
    \left[
    \GAB(\rk,x;s=0)-
    \Smed(sL) 
    \GAB(\rk,x;s)
    \right]
    \,,
    \label{eq:AB-GLV}
    \\
\begin{split}
    x
    \dfrac{\diff^2 I^\text{C}_\text{GLV}}{\diff x \diff(\nk/\thetaC)}
    &=
    \dfrac{4 \alpha_s C_F}{\pi}
    \dfrac{\nk}{\thetaC}
    x^2
    n_0 L \,
    \text{Re}
    \int^1_0 \diff t 
    \int^{\infty}_{0} \dfrac{\diff\nu}{\nu}
    \Big[
    1 - 
    e^{-i \nu^2 (1-t)}
    \Big]
    \\
    &
    \hspace{.20\textwidth}
    \times\Sigma_\text{coh}(t \RC^2 x, 2\RC t \nu\sqrt{x} )
    \Smed(t L)
    G(\kappa/\mu,\sqrt{x}\nu)
    \,,
\end{split}
    \\
    x
    \dfrac{\diff^2 I^\text{Ind}_\text{GLV}}{\diff x \diff(\nk/\thetaC)}
    &=
    \dfrac{4 \alpha_s C_F}{\pi}
    \dfrac{\nk}{\thetaC}
    x^2
    n_0 L
    \int^{\infty}_{0} \dfrac{\diff\nu}{\nu}
    \left[
    1 - 
    \dfrac{\sin(\nu^2)}{\nu^2}
    \right]
    G(\kappa/\mu,\sqrt{x}\nu)
    \,,
    \label{eq:glvhere}
\end{align}
where the functions $H$ and $G$ are defined in eqs.~\eqref{eq:H-init-condition} and~\eqref{eq:broadening-evol-dimensionless}, respectively.

It is also worth noting that the decoherence parameter has not been replaced by its vacuum form, $\Delta_\text{med}\to0$ (i.e. $\Smed\to1$), since it can be easily computed including full resummation of multiple scatterings. We therefore label our results as ``GLV $N=1$ (full $\Delta_\text{med}$)'', and we have checked that the difference with respect to $\Delta_\text{med}\to0$ is minimal.

\section{Results}
\label{sec:results}

The expressions detailed in the previous section, with the exception of those corresponding to the first opacity limit  in subsection~\ref{sec:first-opacity}, are presented for a general linear density of scattering centres $n(s)$. However, in order to numerically solve the final equations, one needs to choose a specific medium model. For simplicity, we consider the QGP `brick' model introduced above. As shown in section \ref{sec:smed}, this choice allows to compute the decoherence parameter $\Delta_\text{med}(t)=1-\Smed(t)$ analytically for both the Yukawa and HTL collision rate models. Nevertheless, we emphasize that the formalism developed in the previous section can be applied to more realistic medium profiles (e.g. expanding media), provided that they satisfy the standard assumptions listed in section \ref{sec:qcd-antenna}.

In the following, we compare results obtained using the Yukawa and HTL scattering rates within both our approach and the GLV approximation. We do not present a comparison with the harmonic oscillator approximation due to the lack of a one-to-one correspondence between its parameters and those of our formalism (and GLV); see, for instance,~\cite{Andres:2020vxs}. Before discussing the results, we recall that the matching between Yukawa and HTL parameters is given in eq.~\eqref{eq:HTL-Yukawa-matching} and follows the procedure in~\cite{Andres:2020vxs}.

\subsection{Decoherence parameter}
\begin{figure}[ht]
\centering
\includegraphics[width=\textwidth]{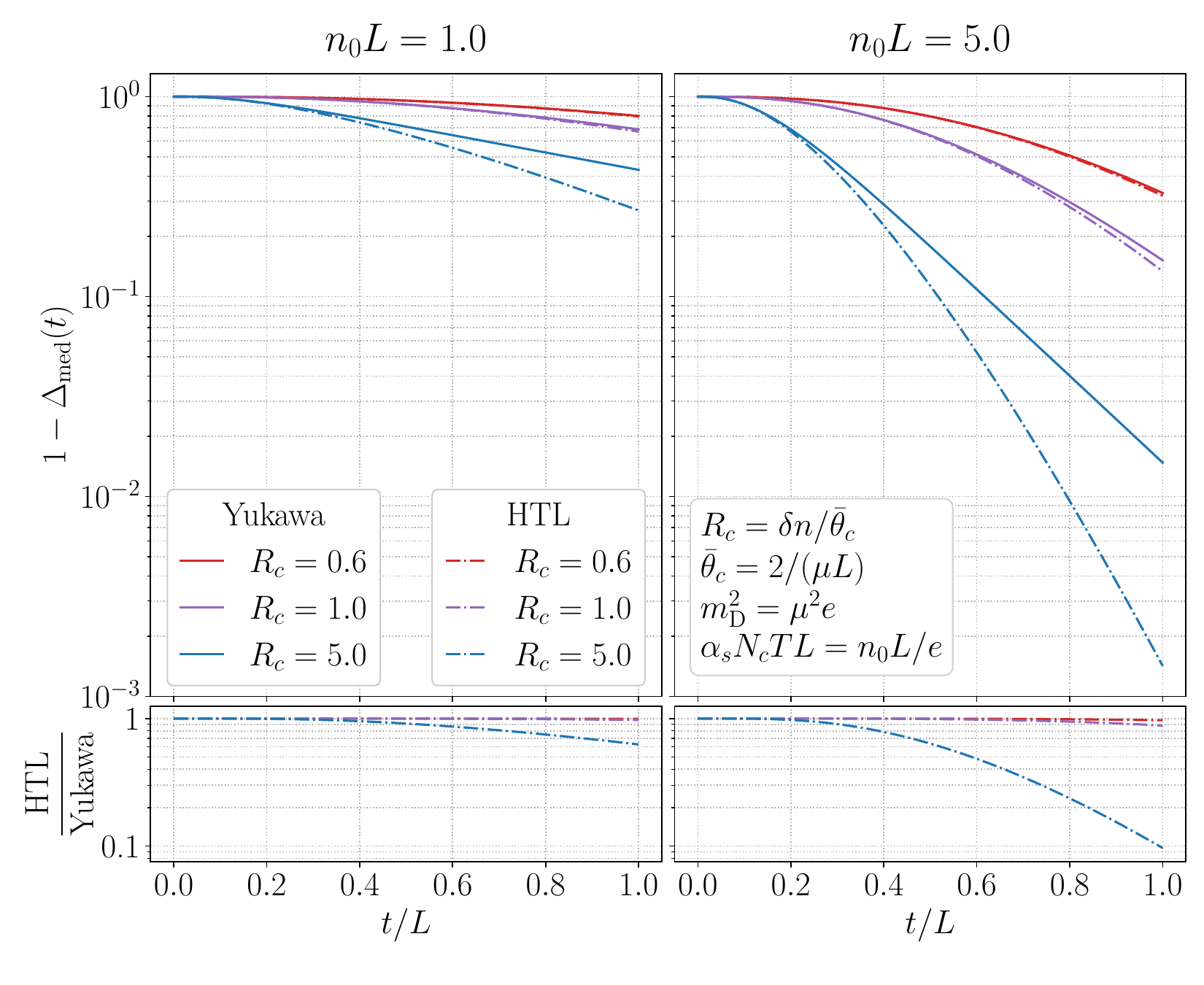}
\vskip -0.5cm
\caption{ $\mathcal{S}_\text{med}=1-\Delta_\text{med}(t)$ of an antenna with opening angle $\dn$ as a function of its propagation time $t$ through a brick with opacity $n_0 L = 1.0$ (left) and $n_0 L = 5.0$ (right). Solid (dash-dotted) curves correspond to a Yukawa (HTL) parton-medium interaction model   for antenna openings corresponding to $\RC\in\{0.6,1.0,5.0\}$ (in \{red, purple, blue\} respectively). The Yukawa and HTL parameters are matched following eq.~\eqref{eq:HTL-Yukawa-matching}.
The bottom panels show the ratio between the HTL and Yukawa scattering rates.}
\label{fig:smed}
\end{figure}

Most contributions to the spectrum are modulated by $\Smed(t) = 1-\Delta_\text{med}(t)$, which quantifies the degree of color decoherence experienced by the antenna prior to gluon emission. In figure~\ref{fig:smed} we show its evolution with time (in units of the medium length $L$) for two fixed values of the opacity $n_0L$ and three of the antenna opening $\RC$.
At a qualitative level, decoherence generally increases with increasing antenna size and medium density. For instance, in the case where the antenna opening is  smaller than the critical angle, i.e $\RC = 0.6$ (red curves), color coherence is largely preserved during propagation  through a dilute medium (left panel) , while it is significantly reduced in a highly opaque medium (right panel).

For narrow antenna configurations ($\dn \lesssim \thetaC$, red and purple), $\Smed$ is rather insensitive to the choice of collision rate model. Their disagreement at the level of $\Smed$  becomes significant only for the widest antenna considered, which features an opening five times larger than the critical angle (blue curve). In this case, the HTL model yields a stronger suppression over time, particularly for antennas traversing a very large/dense medium ($n_0L=5.0$). This behavior contrasts with what was observed for the radiation spectrum off a single color charge~\cite{Andres:2020vxs}, where the details of the interaction become less relevant as the opacity increases. This difference can be understood qualitatively as follows: small antennas remain largely coherent independently of the collision rate,  since the medium cannot efficiently resolve their color structure, so it ``blinds'' them to the details of the potential. In contrast, wide antennas are more sensitive to the infrared structure of the interaction, which is enhanced in the  HTL model.

\subsection{Azimuthally-integrated angular spectrum}

In this section, we aim at discussing coherence effects in the antenna spectrum, together with their evolution with increasing antenna opening angle and opacity.  Results for the fully-resummed azimuthally integrated angular spectrum  both for the Yukawa and HTL collision rates. We further compare them with those obtained in the GLV limit, in which the full form of decoherence parameter, as explained below eq.~\eqref{eq:glvhere}. In particular, we show quark-assigned contribution, and we stress that such decomposition differs from that in~\cite{Mehtar-Tani:2012mfa,Kuzmin:2025fyu}, see subsection~\ref{sec:decomposition}. Further comparisons can be found in appendix~\ref{app:energy-differential-spectra}. 

\begin{figure}[h!]
    \centering
    \includegraphics[width=0.95\textwidth]{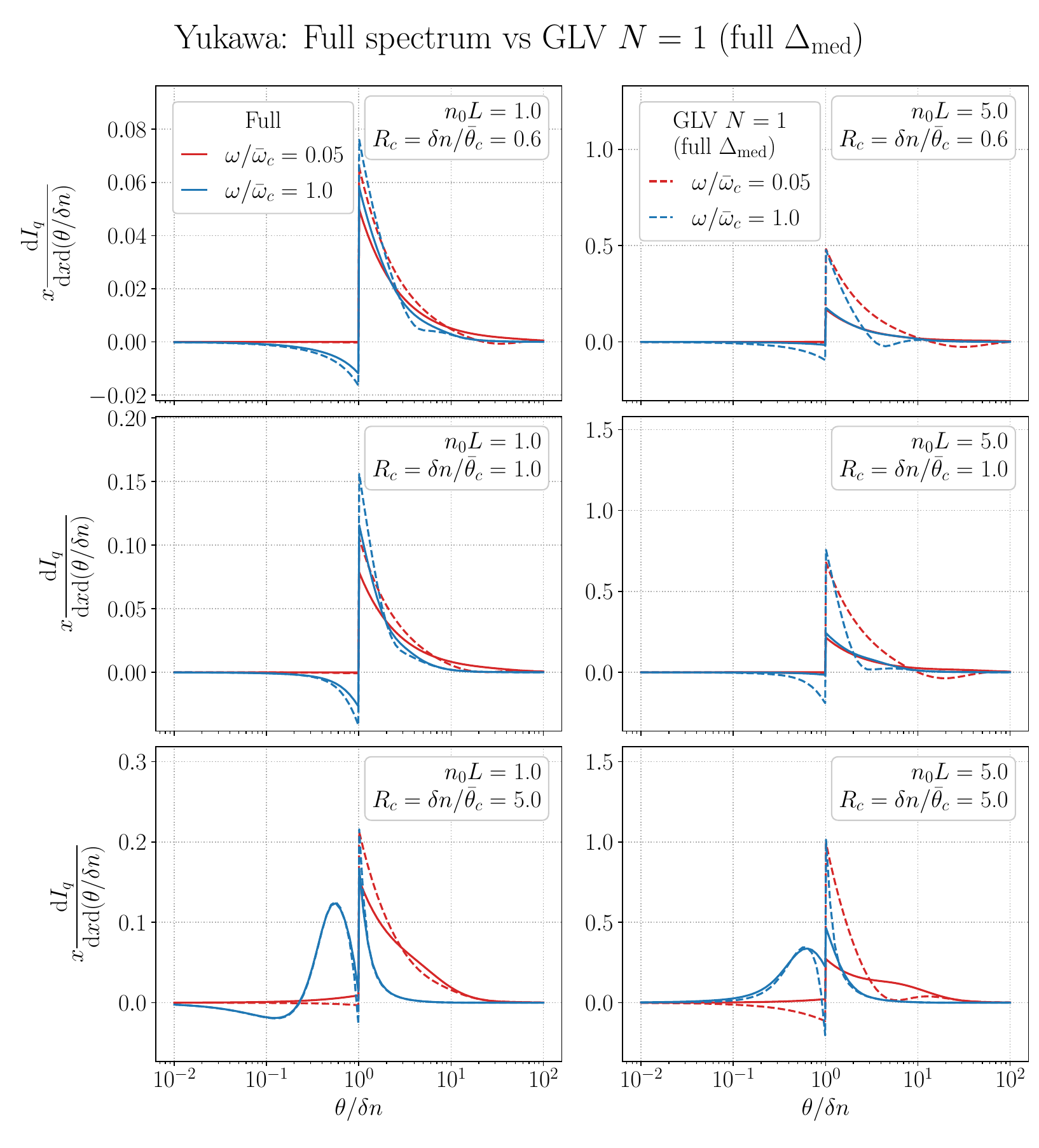}
    \caption{Quark-assigned contribution to the antenna radiation spectrum as a function of the rescaled emission angle $\theta/\delta n$. The left (right) columns correspond   to a medium with $n_0L=1$ ($n_0L=5$), while the rows correspond to antenna openings  $R_c\in\{0.6,1.0,5.0\}$. In each panel, solid (dashed) lines represent the fully-resummed (first opacity) results with the Yukawa scattering rate for fixed values of the rescaled gluon energy, $\omega/\bar{\omega}_c=0.05$ (red), and $\omega/\bar{\omega}_c=1.0$ (blue).}
    \label{fig:angleYuk}
\end{figure}

In figure~\ref{fig:angleYuk} we compare results for the Yukawa scattering model obtained within our approach, labelled as ``Full'' (see solid curves),and under the GLV approximation, labelled as ``GLV $N=1$ (full $\Delta_\text{med}$)'' (see dashed curves). The left column corresponding to a brick with $n_0L=1$ and the right one to a brick with $n_0L=5$. The figure illustrates the transition from coherent ($R_c<1$) to resolved ($R_c>1$) antennas as the opening angle and opacity increase. At a qualitative level, both formalisms agree in their description of in-cone radiation, i.e. for $\theta/\dn < 1$, which  for soft gluons (see red curves), is almost completely independent of the emission angle in all cases. Conversely, more energetic gluons (blue) are sensitive to the individual legs of the antenna and yield a sizeable contribution for $\theta/\dn < 1$. This effect becomes more pronounced as the antenna opening increases.

In the out-of-cone region, i.e. $\theta/\dn >1$, where most of the radiation is emitted, the discrepancies between the Full and GLV are generally larger. As expected, the spectra obtained within the GLV approximation are closer to the full results in a dilute medium (left column). This is particularly true for more energetic gluons (blue curves),whose emission is dominated by a single hard scattering with the medium and is therefore expected to be better described within the $N=1$ opacity limit. However, the agreement between both formalisms  deteriorates noticeably for gluons emitted in an opaque medium (right column). In this regime, the coherent effect of multiple scatterings with the medium becomes increasingly important, causing the GLV approximation to yield a significantly larger spectrum. 

Further, in the case of antennas with smaller opening angles (i.e. $R_c=0.6$, 1.0) we observe the well-known anti-angular ordering property of medium-induced emissions, which restricts the radiation spectrum to the out-of-cone region. It is also worth noting, the presence of local minima of the spectra in the out-of-cone region for the GLV approximation, see, for instance, the blue dashed curve on the top right panel. This feature arises from the interplay between the independent and ($\text{C}$) interference terms, for which the first opacity approximation produces a shifted spectrum relative to the full solution, as already noted in~\cite{Andres:2020vxs}.

As a final remark concerning this figure, we observe that the quark-assigned spectrum develops negative contributions at small angles for both the full solution and the GLV approximation (as well as in the aforementioned out-of-cone ‘dips' featured in the GLV result). A negative in-medium spectrum implies a suppression of radiation with respect to vacuum emissions. As the antenna becomes resolved by the medium ($\Smed\ll 1$), coherent interference effects are progressively suppressed and the spectrum becomes positive, particularly for energetic gluons. The spectrum develops structure near the antenna opening scale due to more localized ``resolved'' emitters. This transition is most visible for high-energy gluons, which probe shorter transverse scales and are therefore more sensitive to the individual legs of the antenna, while softer gluons are more broadly distributed.

\begin{figure}[h!]
    \centering
    \includegraphics[width=0.95\textwidth]{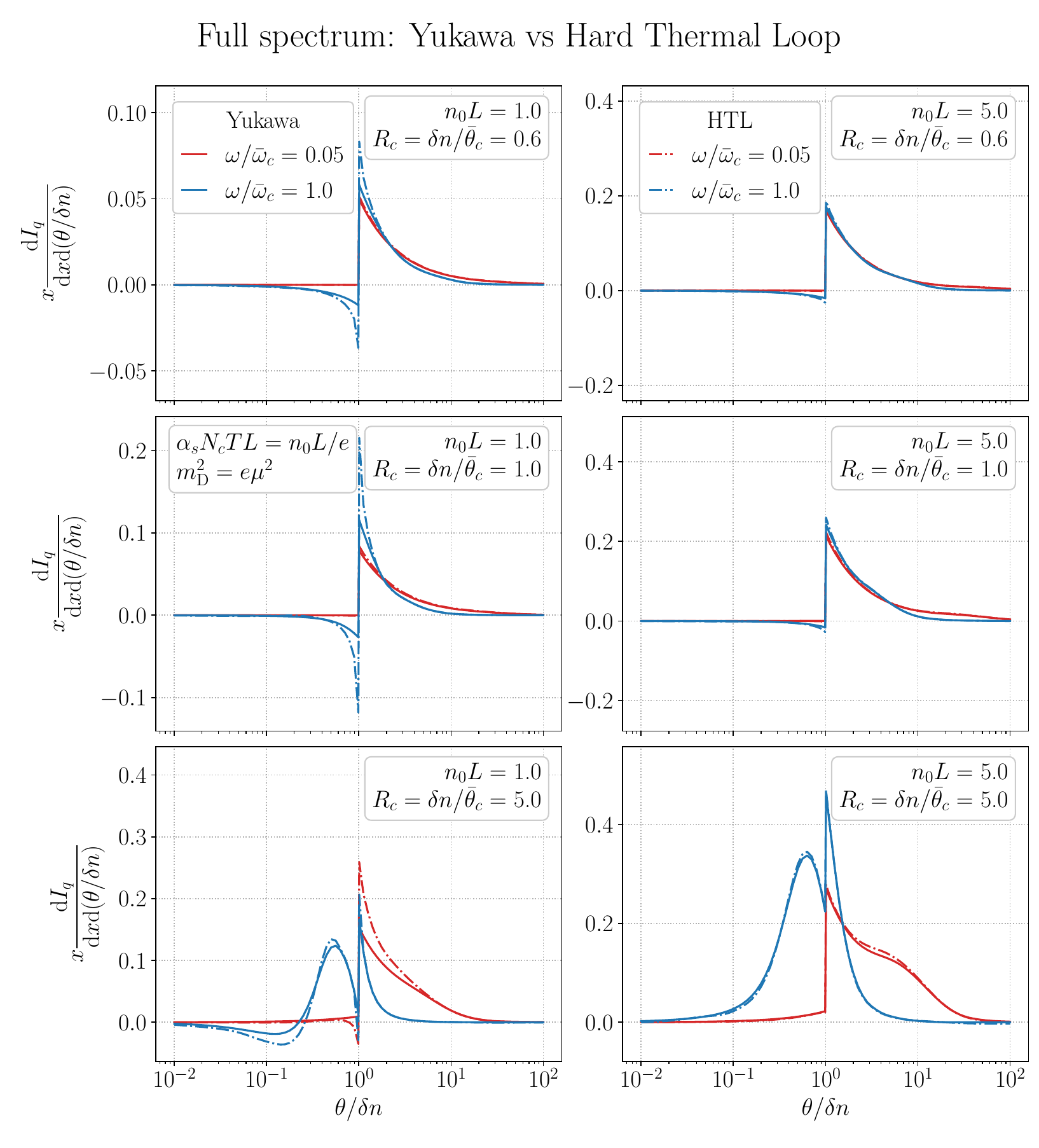}
    \caption{Quark-assigned contribution to the fully resummed antenna radiation spectrum as a function of the rescaled emission angle $\theta/\delta n$. The left (right) columns correspond   to a medium with $n_0L=1$ ($n_0L=5$), while the rows correspond to antenna openings $R_c\in\{0.6,1.0,5.0\}$. In each panel, solid (dash-dotted) lines represent the fully-resummed result for the Yukawa  (HTL) scattering rates for fixed values of the rescaled gluon energy, $\omega/\bar{\omega}_c=0.05$ (red), and $\omega/\bar{\omega}_c=1.0$ (blue).}
    \label{fig:angleYukHTL}
\end{figure}

We now turn to figure~\ref{fig:angleYukHTL}, where we compare results obtained within the fully resummed formalism developed in this manuscript for the Yukawa (solid curves) and HTL (dash-dotted curves) parton-medium interaction models. Overall, the agreement is significantly better than between the full Yukawa spectrum and the GLV  approximation, indicating that resumming multiple scatterings has a larger impact than the precise rate choice -- at least for the azimuthally integrated spectra shown here. In contrast with the behaviour discussed in the previous section, the agreement is almost perfect in the opaque medium case (right column) for all values of $R_c$, and across the full range of emission angles shown. Discrepancies become notable only when considering a dilute medium (left column) and  for resolved or intermediate antennas. This observation is consistent with the conclusions of~\cite{Andres:2020vxs}, which suggest that the details of the medium interaction become less relevant as the opacity increases.  Notably, the similarity of both results at large opacities persists even when the decoherence parameter becomes very different for the two models, see figure~{\ref{fig:smed}}. This can be understood by noting that, for a wide antenna in an opaque medium, decoherence effects become so strong that the spectrum becomes dominated by independent emissions and the B interference term. Since these terms do not depend on the decoherence parameter, the resulting spectra are fairly insensitive to the details of the scattering rate.

\section{Conclusions}
\label{sec:conclusion}

In this work, we have developed a framework to compute the soft-gluon emission spectrum of an in-medium antenna, fully accounting for multiple scatterings without additional approximations. We use Dyson-type integral equations to evaluate this azimuthally integrated emission spectrum off a quark–antiquark pair, extending the method of~\cite{Andres:2020vxs}, originally developed for a single emitter. This approach enables a complete resummation of medium interactions and can be applied to realistic parton-interaction models without further simplifying assumptions (such as the harmonic oscillator).

The resulting expressions are solved numerically within a Yukawa interaction model to obtain the fully differential angular emission spectrum of the antenna. These results provide a direct illustration of the transition from coherent to resolved antennas, as the opening angle and medium opacity increase. As the antenna becomes resolved by the medium, coherent interference effects are progressively reduced relative to the independent-emission component, and the emitted gluon approaches the radiation pattern expected from independently radiating color charges. A comparison with the GLV approximation reveals significant differences, particularly in opaque media and in the out-of-cone region, where multiple scatterings play an essential role. In these regimes, the GLV approach yields a significantly harder spectrum, while agreement is recovered for sufficiently energetic gluons whose emission is dominated by a single hard scattering with the medium.

We further evaluate the fully resummed angular antenna spectrum using the hard-thermal-loop description of parton-medium interactions. To assess the impact of the interaction model, we confront these results with those obtained in the Yukawa framework, with the parameters of the two models chosen to reproduce the same leading-order asymptotic behavior at large gluon energies and transverse momenta. Overall, we find a much better agreement than between the full Yukawa and GLV results, with discrepancies decreasing as the medium opacity grows. 

We emphasize that a main motivation for this work is the well-known results from single-emitter case that neither the resummation of multiple scatterings under the harmonic oscillator approximation nor the truncation at first order in opacity provides an accurate spectrum over the full kinematic range of the emitted gluon \mbox{\cite{Andres:2020kfg}}. In this context, the similarity between the fully resummed antenna spectra obtained using Yukawa and HTL interaction rates obtained in this manuscript suggests that the resummation of multiple scatterings is the dominant ingredient for obtaining an accurate antenna spectrum, while the precise form of the interaction rate plays a comparatively subleading role, provided it exhibits the correct Coulomb-like behavior at large momentum transfers (a property not satisfied by the harmonic oscillator approximation). In fact, we observe that even in configurations where the Yukawa and HTL models predict noticeably different decoherence rates, the resulting radiation spectra remain remarkably similar once multiple scatterings are fully resummed. This observation suggests that coherence-driven observables may be more robust against the details of the microscopic interaction model, provided that the effects of multiple scatterings are consistently included. A natural extension would be to replace the analytic Yukawa and HTL interaction kernels by collision kernels obtained from first-principles calculations, including QCD kinetic-theory simulations and, where available, non-perturbative determinations based on lattice QCD~\cite{Moore:2021jwe,BarreraCabodevila:2025ogv,Lindenbauer:2025ctw}.

In conclusion, we have developed a method to compute the in-medium gluon spectrum off a QCD antenna for arbitrary scattering models, while fully resumming multiple interactions with the medium. Beyond its methodological interest in enabling fully resummed calculations for arbitrary collision rates, this framework provides a new tool to investigate the dynamics of color coherence in QCD media. While the present work focuses on azimuthally integrated spectra in a static medium, the formalism is readily applicable to more realistic scenarios and opens the way for future phenomenological studies of coherence effects in jet observables. In particular, medium-induced decoherence is often modelled in phenomenological semi-analytical studies within the harmonic oscillator approximation. Our results provide a framework to go beyond this limit and quantify the impact of more realistic interaction models on color decoherence phenomena.

\section*{Acknowledgements}

The authors would like to thank João Barata, Florian Lindenbauer, João Silva and Konrad Tywoniuk for helpful discussions. 

This research was supported by European Research Council projects ERC-2018-ADG-835105 YoctoLHC and QGPthroughEECs, grant agreement No.~101164102. Views and opinions expressed are however those of the authors only and do not necessarily reflect those of the European Union or the European Research Council. Neither the European Union nor the granting authority can be held responsible for them. This work was also supported by Xunta de Galicia (CIGUS Network of Research Centres) European Union ERDF, by the Spanish Research State Agency under projects PID2023152762\-NB--I00 and CEX2023001318-M financed by MICIU/AEI/10.13039/501100011033, by the Portuguese Fundação para a Ciência e a Tecnologia (FCT), under ERC-PT A-Projects ‘Unveiling’, financed by PRR, NextGenerationEU. This project received funding from the Ecole Polytechnique Foundation. 

LA acknowledge support by FCT under contract 2021.03209.CEECIND and AC acknowledges FCT support under contract PRT/BD/154190/2022. 

\appendix

\section{Azimuthal averages}
\label{app:averages-details}

\subsection{Averaging the scattering rates}

The main text relies on the following angular moments for the scattering rate,
\begin{align}
    M_n(k,q;\mu)
    =
    \int^{2\pi}_0 \dfrac{\diff\varphi}{2\pi}
    \dfrac{V(k,q,\cos\varphi;\mu)}{4\pi}
    \cos(n\varphi)
    \,,
\label{eq:rate-harmonics}
\end{align}
\noindent where $\varphi$ is the angle between the momenta $\bm{k}$ and $\bm{q}$, i.e. $(\bm{k-q})^2  = k^2+q^2-2kq\cos\varphi$.

\subsubsection{Yukawa scattering rate}

For the Yukawa (or Gyulassy-Wang) model, we have the following scattering rate:
\begin{align}
    V(\bm{k-q}) 
    = \dfrac{8\pi\mu^2}{((\bm{k-q})^2+\mu^2)^2}
    \,,
\end{align}
\noindent for which the Fourier coefficients in eq.~\eqref{eq:rate-harmonics} are:
\begin{align}
    M_n(k,q;\mu) &= 
    \dfrac{\mu^2}{4 k^2 q^2}
    \dfrac{
    (n+1)z(k,q,\mu)^{n-1}
    -
    (n-1)z(k,q,\mu)^{n+1}
    }{
    [f(k,q,\mu)^2-1]^{3/2}
    } \,,
    \\
    f(k,q,\mu) &=
    \dfrac{k^2+q^2+\mu^2}{2kq} \,,
    \\
    z(k,q,\mu) &=
    f(k,q,\mu) - \sqrt{f(k,q,\mu)^2-1}
    \label{eq:z-aux}
    \,,
\end{align}
\noindent having defined the auxiliary functions $f(k,q,\mu)$ and $z(k,q,\mu)$. For convenience, we list the zeroth and first moments:
\begin{align}
    M_0(k,q;\mu)
    &=
    2\mu^2
    \dfrac{k^2+q^2+\mu^2}{[(k^2+q^2+\mu^2)^2 - 4k^2q^2]^{3/2}}
    \,,
    \\
    M_1(k,q;\mu)
    &=
    2\mu^2
    \dfrac{2kq}{[(k^2+q^2+\mu^2)^2 - 4k^2q^2]^{3/2}}
    \,.
\end{align}

Further, the initial condition in eq.~\eqref{eq:broadening-evol-dimensionless} reads:
\begin{align}
    G\left(\frac{k}{\mu}=K, \frac{q}{\mu}=Q\right)
    =
    \dfrac{2(K^2-Q^2+1)}{
    [(K^2+Q^2+1)^2 - 4K^2Q^2]^{3/2}
    }
    \,.
\end{align}

\subsubsection{HTL scattering rate}

For the Hard Thermal Loop model, we use the scattering rate:
\begin{align}
    V(\bm{k-q}) =
    \dfrac{8\pi\md^2}{(\bm{k-q})^2((\bm{k-q})^2 + \md^2)}
    \,,
\end{align}
\noindent with the following coefficients:
\begin{align}
    M_n(k,q;\md) &=
    2
    \dfrac{z(k,q,0)^n}{|k^2-q^2|}
    -
    2
    \dfrac{z(k,q,\md)^n}{\sqrt{(k^2+q^2+\md^2)^2 - 4k^2q^2}}
    \,,
\end{align}
\noindent using the same auxiliary function $z(k,q,\md)$ as in eq.~\eqref{eq:z-aux}. For convenience, the zeroth and first scattering moments are:
\begin{align}
    M_0(k,q;\md)
    &=
    \dfrac{2}{|k^2-q^2|}
    -
    \dfrac{2}{\sqrt{(k^2+q^2+\md^2)^2 - 4k^2q^2}}
    \,,
    \\
    M_1(k,q;\md)
    &=
    \dfrac{1}{kq}
    \left[
    \dfrac{k^2+q^2}{|k^2-q^2|}
    -
    \dfrac{k^2+q^2+\md^2}{\sqrt{(k^2+q^2+\md^2)^2 - 4k^2q^2}}
    \right]
    \,.
\end{align}

We finally note that, despite the singularities in the first term of $M_0$ and $M_1$, the initial condition in eq.~\eqref{eq:broadening-evol-dimensionless} is regular, and reads:
\begin{align}
    G\left(\frac{k}{\md}=K, \frac{q}{\md}=Q\right)
    =
    \dfrac{1}{K^2}
    \left[
    \text{sgn}\left(K^2-Q^2\right)
    +
    \dfrac{-K^2+Q^2+1}{
    \sqrt{(K^2+Q^2+1)^2 - 4K^2Q^2}
    }
    \right]
    \,.
\end{align}

\subsection{Averaging the spectrum}

Here we provide some details regarding the angular integrals in eqs.~\eqref{eq:azimuthal-average-psiAB}~and~\eqref{eq:azimuthal-average-Psi}. The integral in question is:
\begin{align}
\begin{split}    
    \tilde{\Psi}(\kappa,\omega;s)
    =&
    \int^{2\pi}_{0} \dfrac{\diff\varphi_\kappa}{2\pi}
    \int^{2\pi}_{0} \dfrac{\diff\varphi_q}{2\pi}
    \int^{\infty}_{0} \diff q q
    \dfrac{V(\kappa,q,\cos\varphi_{\kappa q};\mu)}{4\pi}
    \\
    &\times
    \left[
    \dn^2 \mathcal{J}^\text{vac}
    (\bm{\kappa}, \bm{\kappabar})
    -
    \dn^2 \mathcal{J}^\text{vac}
    (\bm{q}, \bm{\bar{q}})
    \right]
    \times
    \text{Re }
    e^{i s \tfrac{ \bm{\bar{q}}^2 - \bm{q}^2 }{2\omega}}
    \,,
\end{split}
\end{align}
where, as in the main text, $\varphi_q$ is the azimuthal angle for the transverse momentum $\bm{q}$, and $\varphi_{\kappa q} \equiv \varphi_\kappa - \varphi_q$. Here, we intend to perform the angular integrals analytically, leaving the $q$ integration to be performed numerically.

Firstly, the terms proportional to the vacuum interference, e.g. $\mathcal{J}^\text{vac}(\bm{q}, \bm{\bar{q}})$, can be written in terms of the azimuthal angle $\varphi_q$ and the rescaled emission angle $r_q = q/(\omega \dn)$:
\begin{align}
    \dn^2 \mathcal{J}^\text{vac} (\bm{q}, \bm{\bar{q}})
    =
    \dfrac{4}{r^2_q} D(\cos\varphi_q; r_q)
    \,,
\end{align}
\noindent using the auxiliary function $D(\cos\varphi; t)$ with the following cosine series representation (from the generating function for Chebyshev polynomials):
\begin{align}
    D(\cos\varphi;t) 
    &= 
    \dfrac{t-\cos\varphi}{t^2-2t\cos\varphi+1} 
    = 
    \sum^{+\infty}_{n=-\infty} 
    D_n(t) 
    \cos(n\varphi)
    \,,
    \\
    D_0(t) 
    &=
    \Theta(|t|-1) 
    \,,\\
    D_{n}(t)
    &= 
    \dfrac{
    \Theta(|t|-1) t^{-|n|}
    -
    \Theta(1-|t|) t^{+|n|}    
    }{2}
    \qquad  {\rm for }\qquad n\neq0
    \,,
\end{align}
such that $D_{-n}(t) = D_{n}(t)$, and where $\Theta(x)$ is the usual Heaviside step function, where we adopt the convention $\Theta(0)=1/2$.

Secondly, the complex phase in the integral can be simplified using the generating function for Bessel functions:
\begin{align}
    e^{-iz\cos\varphi}=
    J_0(z)
    +
    2\sum^{\infty}_{n=1}
    e^{-in\pi/2}
    J_n(z) \cos(n\varphi)
    \,,
\end{align}
\noindent such that one obtains:
\begin{align}
    \text{Re }
    e^{i s \tfrac{ \bm{\bar{q}}^2 - \bm{q}^2 }{2\omega}}
    =&
    \text{ Re}
    \exp\left(is \frac{\omega\dn^2}{2}\right)
    \exp\left(-i \dn s q \cos\varphi_q\right)
    =
    \\
\begin{split}        
    =&
    \cos\left(s \frac{\omega\dn^2}{2}\right)
    \times
    \left[
    J_0(\dn s q)
    +
    2\sum^{\infty}_{j=1} 
    (-1)^j
    J_{2j}(\dn s q)
    \cos(2j\varphi_q)
    \right]
    +
    \\
    &+
    \sin\left(s \frac{\omega\dn^2}{2}\right)    
    \times
    2\sum^{\infty}_{j=1} 
    (-1)^j
    J_{2j-1}(\dn s q)
    \cos[(2j-1)\varphi_q]
    \,.
\end{split}
\end{align}

Thirdly, we consider the scattering rate, which can be expanded in the cosine series:
\begin{align}
    \dfrac{V(\kappa,q,\cos\varphi_{\kappa q}; \mu)}{4\pi}
    &=
    \sum^{+\infty}_{n=-\infty}
    M_n(\kappa,q;\mu)
    \cos(n(\varphi_\kappa - \varphi_q))
    =
    \\
    &=
    \sum^{+\infty}_{n=-\infty}
    M_n(\kappa,q;\mu)
    \left[
    \cos(n\varphi_\kappa)
    \cos(n\varphi_q)
    +
    \sin(n\varphi_\kappa)
    \sin(n\varphi_q)
    \right]
    \,,
\end{align}
\noindent where the sine terms vanish under integration against functions of $\cos\varphi_\kappa$ and $\cos\varphi_q$.

Putting these three results together, the angular integrals can be performed, since they are all of the form:
\begin{align}
    \int^{2\pi}_{0} \dfrac{\diff\varphi}{2\pi}
    \cos(x\varphi)\cos(y\varphi)\cos(z\varphi)
    =
    \dfrac{
    \delta_{x+y+z=0}
    +\delta_{x-y+z=0}
    +\delta_{x+y-z=0}
    +\delta_{x-y-z=0}
    }{4}
    \,.
\end{align}

After performing all integrals and sums, appropriately rescaling all variables, as well as identifying the integrals to be performed numerically, we find the results presented in eqs.~\eqref{eq:H-init-condition}~and~\eqref{eq:I-integrals-def}.

\section{Energy differential spectra}
\label{app:energy-differential-spectra}

\begin{figure}[h!]
    \centering
    \includegraphics[width=0.99\textwidth]{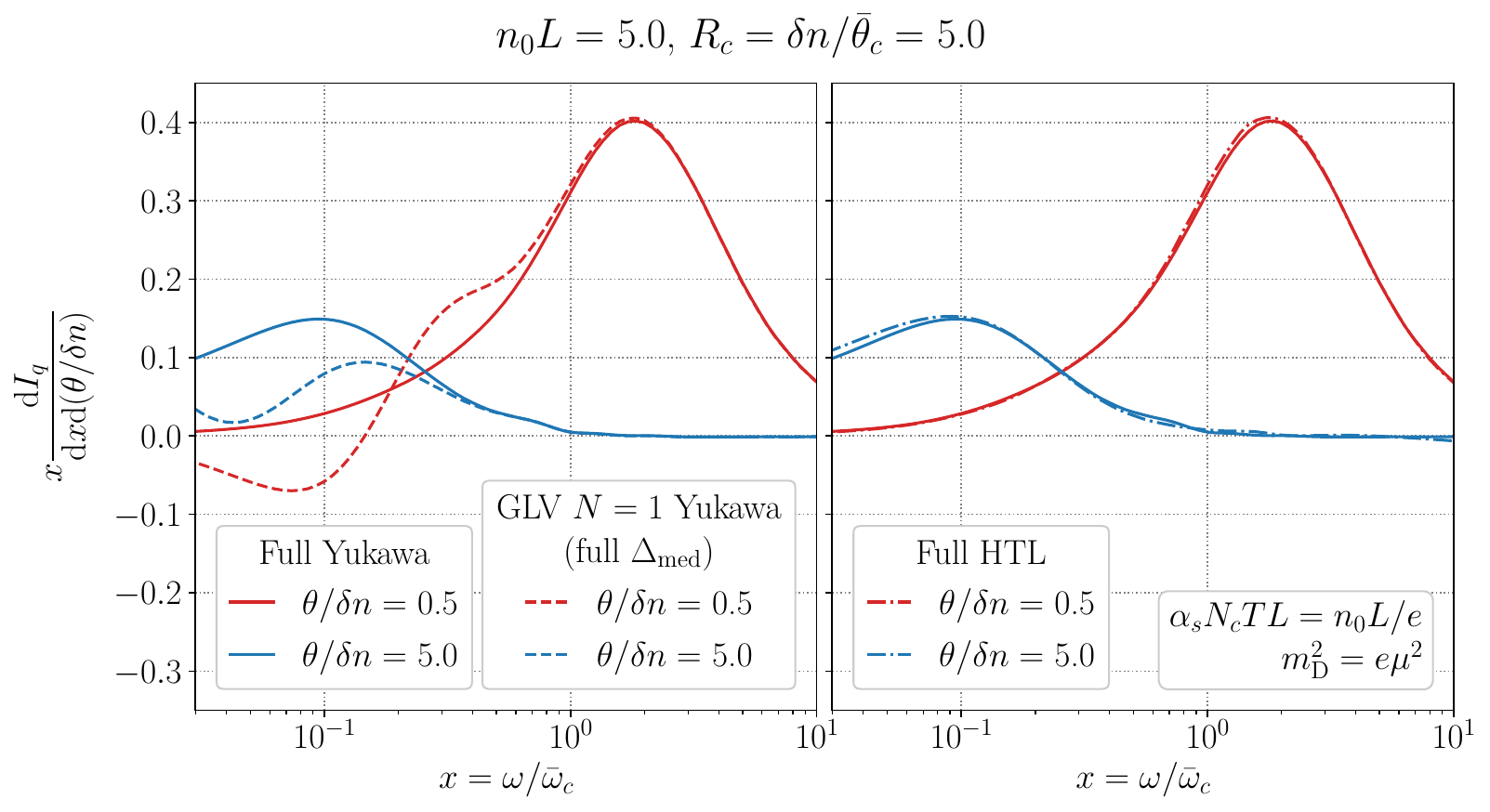}
    \caption{Quark-assigned contribution to the gluon energy spectrum of an antenna with opening angle $R_c=5.0$ inside a medium of opacity $n_0L=5.0$ as a function of the rescaled gluon energy $\omega/\bar{\omega}_c$. The full result for the Yukawa scattering rate (solid lines) is compared with the GLV first opacity approximation (dashed lines) in the left panel, and with the full result for the HTL scattering rate (dot-dashed lines) in the right panel. The red and blue curves correspond to fixed values of the rescaled emission angle $\theta/\delta n\in\{0.5,5.0\}$, respectively.}
    \label{fig:Edist}
\end{figure}

To further clarify the comparison between the Yukawa and HTL models and the GLV approximation, in figure~\ref{fig:Edist} we show the gluon energy spectra for two values of emission angles (in and out of cone), for a dense medium and an antenna with large opening angle. We observe significant differences concentrated in the soft part of the spectrum ($\omega<\bar{\omega}_c$) between the full and GLV $N=1$ results for the Yukawa model (see left panel). Although the GLV approximation typically overestimates the spectrum relative to the full calculation (due to the absence of coherence effects), the discrepancies shown below $x\sim0.2$ precisely correspond to the two regions of the $n_0L=5.0$, $R_c=5.0$ soft gluon spectrum where this is not the case. For $\theta/\delta n=5.0$ (blue curves), this corresponds to the out-of-cone minimum featured in the GLV spectrum, which is caused by the interplay between direct and interference terms. For the $\theta/\delta n=0.5$ case (red curves), the discrepancy corresponds to the region where the GLV spectrum becomes negative. Both of these features were discussed in the main text of the manuscript. Conversely, the agreement between the Yukawa and HTL models is almost perfect for both emission angles and throughout the whole energy range represented. More generally, we also observe that, as expected, soft (hard) gluons emission happen dominantly at large (small) angles, in agreement with the previous discussions.

\bibliography{biblio}
\bibliographystyle{JHEP}

\end{document}